\documentclass[12pt,reqno,a4paper]{article}
\usepackage{amsmath,amsfonts,amssymb,graphicx,epsfig}
\numberwithin{equation}{section}

\newcommand{\nc}{\newcommand}
\nc{\rnc}{\renewcommand}

\nc{\be}{\begin{equation}}
\nc{\round}[1]{\left(#1\right)}
\nc{\squareb}[1]{\left[#1\right]}
\nc{\curly}[1]{\left\{#1\right\}}
\nc{\abs}[1]{\left|#1\right|}
\nc{\pointy}[1]{\left\langle #1 \right\rangle}
\nc{\lsum}[3]{\sum_{#1=#2}^{#3}}

\nc{\Wt}[4]{W
\begin{pmatrix}
#4 & #3\\
#1 & #2
\end{pmatrix}}

\nc{\Kt}[3]{K
\begin{pmatrix}
#2 &
\begin{matrix}
#3\\
#1
\end{matrix}
\end{pmatrix}}

\nc{\Db}{D_{\text{b}}}
\nc{\Ds}{D_{\text{s}}}
\nc{\Df}{D_{\text{f}}}
\nc{\Tb}{T_{\text{b}}}
\nc{\ks}{\kappa_{\text{s}}}
\nc{\kb}{\kappa_{\text{b}}}
\nc{\fsing}{f_{\text{sing}}}
\nc{\fsinglr}{\fsing^{\text{L},\text{R}}}
\nc{\Cb}{C_{\text{b}}}
\nc{\Cs}{C_{\text{s}}}
\nc{\als}{\alpha_{\text{s}}}
\nc{\fb}{f_{\text{b}}}
\nc{\fs}{f_{\text{s}}}

\nc{\xil}{\xi_{\text{L}}}
\nc{\xir}{\xi_{\text{R}}}
\nc{\al}{a_{\text{L}}}
\nc{\ar}{a_{\text{R}}}
\nc{\kslr}{\ks^{\text{L},\text{R}}}
\nc{\epsl}{\eps^{\text{L}}}
\nc{\epsr}{\eps^{\text{R}}}
\nc{\epslr}{\eps^{\text{L},\text{R}}}

\nc{\thf}{\vartheta_1}
\nc{\thfr}{\vartheta_4}

\nc{\vm}{\boldsymbol{m}}
\nc{\vn}{\boldsymbol{n}}
\nc{\ve}{\boldsymbol{e}}
\nc{\vI}{\mathcal{I}}
\nc{\A}{\mathrm{A}}
\nc{\one}{$1$\nobreakdash}
\nc{\two}{$2$\nobreakdash}
\def\Re{\mbox{Re}}
\def\Im{\mbox{Im}}
\def\half {\mbox{$\textstyle {1 \over 2}$}}

\nc{\lam}{\lambda}
\nc{\Lam}{\Lambda}
\nc{\Om}{\Omega}
\nc{\gam}{\gamma}
\nc{\eps}{\epsilon}
\nc{\veps}{\varepsilon}
\nc{\kap}{\kappa}
\nc{\m}{\mu}

\nc{\D}{\boldsymbol{D}}
\nc{\av}{\boldsymbol{a}}
\nc{\bv}{\boldsymbol{b}}
\nc{\ttt}{\boldsymbol{t}}
\nc{\I}{\boldsymbol{I}}

\nc{\sm}[1]{{\scriptstyle #1}}
\nc{\ssm}[1]{{\scriptscriptstyle #1}}
\nc{\spos}[2]{\makebox(0,0)[#1]{$\sm{#2}$}}
\nc{\dl}[3]{\put(#1,#2){\makebox(#3,0){\dotfill}}}
\rnc{\d}[2]{\put(#1,#2){\spos{}{\bullet}}}
\nc{\dd}[3]{\multiput(#1,#2)(0,1){#3}{\spos{}{\bullet}}}

\def\Mult#1#2#3{{\genfrac{[}{]}{0pt}{0}{#1}{#2}}_{#3}}
\def\u{\epsilon}
\def\b{\beta}

\def\h{\vartheta}
\def\xR{\xi_R}
\def\xL{\xi_L}

\def\r2{\rho_2}
\def\({\biggl(}
\def\){\biggr)}

\def\l{\lambda}

\rnc{\title}[1]{{\Large\bf\mbox{}\\\medskip#1\bigskip\medskip\\}}
\rnc{\author}[1]{{\large #1\smallskip\\}}
\nc{\address}[1]{{\em #1\medskip\\}}

\def\gauss#1#2{\mbox{\small $\left[#1\atop #2\right]$}}
\def\floor#1{\lfloor #1\rfloor}

\begin{document}

\begin{center}

\title{Excited TBA Equations II:\\
Massless Flow from Tricritical to Critical Ising Model}

\author{
Paul A. Pearce,\!\!
\footnote{
P.Pearce@ms.unimelb.edu.au}\;
Leung Chim\!\!
\footnote{Current address: DSTO, Adelaide; 
Leung.Chim@dsto.defence.gov.au}
}
\address{Department of Mathematics and Statistics\\
University of Melbourne, 
Parkville, Victoria 3010, Australia}
\author{Changrim Ahn\!\!
\footnote{ahn@dante.ewha.ac.kr}}
\address{Department of Physics\\
Ewha Womans University, Seoul 120-750, Korea}


\begin{abstract}
\noindent
We consider the massless tricritical Ising model ${\cal M}(4,5)$ perturbed by
the thermal operator $\varphi_{1,3}$ in a cylindrical geometry and apply
integrable boundary conditions, labelled by the Kac labels $(r,s)$, 
that are natural off-critical perturbations of known conformal boundary conditions. 
We derive massless thermodynamic Bethe ansatz (TBA) equations for all excitations by 
solving, in the continuum scaling limit, the TBA functional equation satisfied by the double-row 
transfer matrices of the $A_4$ lattice model of Andrews, Baxter and Forrester (ABF) in 
Regime~IV. 
The resulting TBA equations describe the massless renormalization group flow from the
tricritical to critical Ising model. As in the massive case of Part~I, the excitations are
completely classified in terms of $(\vm,\vn)$ systems but the string content changes
by one of three mechanisms along the flow. Using generalized $q$-Vandemonde identities, we show
that this leads to a flow from tricritical to critical Ising characters. The excited TBA
equations are solved numerically to follow the continuous flows from the UV to the IR
conformal fixed points.
\end{abstract}
\end{center}

\section{Introduction}

In integrable Quantum Field Theory (QFT), the Thermodynamic Bethe Ansatz (TBA)~\cite{YangYang,Zam} continues to be an important method for the study of massive and massless Renormalization Group (RG) flows including the study of excited states~\cite{Martins,Fendley,DoreyTateo}. Moreover the Tricritical Ising Model (TIM), which is the simplest member ${\cal M}(4,5)$ of the unitary minimal series~\cite{BPZ84} beyond the critical Ising model ${\cal M}(3,4)$, remains a rich example for studying both thermal and boundary flows~\cite{Lesage,FPR02,Nepo02,NepoAhn02}.

In Part~I of this series (hereafter referred to as PCAI~\cite{PCAI}) we considered the
massive tricritical Ising model ${\cal M}(4,5)$ perturbed by the thermal operator $\varphi_{1,3}$
in a cylindrical geometry and systematically derived the TBA equations for all excitations by
using a lattice approach. More specifically this was achieved by solving, in the continuum
scaling limit, the TBA functional equation satisfied by the double-row transfer matrices of the
$A_4$ lattice model of Andrews, Baxter and Forrester (ABF) in  Regime~III~\cite{ABF84}. 
In this paper we turn
our attention to the massless tricritical Ising model which is obtained as the continuum scaling
limit of the $A_4$ lattice model in Regime~IV. Our goal is to systematically derive and study
the massless TBA equations which describe the renormalization group (RG) flow from the
tricritical ${\cal M}(4,5)$ to critical ${\cal M}(3,4)$ Ising model. We apply the methods
developed in PCAI and use the concepts and notations introduced in that paper without further
elaboration.

The layout of the paper is as follows. In Section~2 we discuss the classification of excitations using $(\vm,\vn)$ systems including a description of the three mechanisms by which the string contents change along the flow. In Section~3 we derive in detail the massless TBA equations in the $(r,s)=(1,1)$ vacuum sector. We do not discuss the very similar derivation of the TBA equations in the other 5 sectors. The numerical
solution of the TBA equations to yield continuous flows is discussed in Section~4. We finish in Section~5 with a brief discussion.

\section{Classification of Excitations and Flows}

For small perturbations, the scaling limit of the excitations in Regime~IV of the $A_4$ lattice model are classified by precisely the same $(\vm,\vn)$ systems~\cite{Melzer,Berk94} as at the conformal critical point~\cite{OPW97} and throughout the massive Regime~III~\cite{PCAI}. Unlike
the massive case, however, we find that in the massless regime the string content can change by one of three mechanisms along the flow. This was first observed empirically by  direct numerical diagonalization of a sequence of finite-size transfer matrices approaching the scaling limit
\begin{equation}
\mu=\frac{mR}{4}=\lim_{N\to\infty,\, t\to 0} N |t|^\nu,\qquad \nu=5/4
\label{scalinglimit}
\end{equation}
for modest values of the system size $N$ and $0\le mR< 5$ and is confirmed by our numerical solutions of the TBA equations. 
Here $\mu$ measures the perturbation strength and $t$ is the departure-from-criticality variable. The mass $m$ and continuum length scale $R$ usually occur in the single combination $mR$.
Notice that in the finite-size scaling we use the Regime~III correlation length exponent $\nu=5/4$ even though the actual~\cite{BaxP83}  correlation length exponent in Regime~IV is $\nu'=5/2$.

Let us consider the vacuum sector with boundary condition
$(r,s)=(1,1)$. The excitation energies are given by the scaling limit of the eigenvalues of the
double-row transfer matrix $\D(u)$, or equivalently the normalized transfer matrix $\ttt(u)$,
where $u$ is the spectral parameter. The two relevant analyticity strips in the complex $u$-plane
are respectively 
\begin{equation}
-\frac{\lambda}{2}<\Re(u)<\frac{3\lambda}{2},\qquad 2\lambda<\Re(u)<4\lambda
\end{equation}
where $\lambda=\pi/5$ is the crossing parameter. The excitations are classified by their string
contents 
\begin{equation}
\begin{split}
m_i&=\text{\{number of \one-strings in strip $i=1,2$\}}\\
n_i&=\text{\{number of \two-strings in strip $i=1,2$\}}
\end{split}
\label{stringdefs}
\end{equation}
At the conformal critical point ($mR=0$) the string contents satisfy the $(\vm,\vn)$ system
\begin{equation}
\vm + \vn = \frac{1}{2}(N\ve_{1}+\vI\vm)
\label{mn11}
\end{equation}
where $m_1, m_2$ and $N$ are even, $\vm=(m_{1},m_{2})$, $\vn=(n_{1},n_{2})$, $\ve_{1}=(1,0)$, and
$\vI$ is the $\A_{2}$ incidence matrix with entries
$\vI_{j,k}=\delta_{|j-k|,1}$. For the leading excitations $m_1, m_2, n_2$ are finite but 
$n_1\sim N/2$ as $N\to\infty$. 

As explained in PCAI, an excitation with string content $(\vm,\vn)$ is uniquely labelled by a set of quantum numbers
\begin{equation}
I=(I^{(1)}|I^{(2)})=(I_1^{(1)},I_2^{(1)},\ldots,I_{m_1}^{(1)}\;|\;I_1^{(2)},I_2^{(2)},
\ldots,I_{m_2}^{(2)})
\label{quantumnos}
\end{equation}
where the integers $I_j^{(i)}\in\{0,1,2,\ldots\}$ with $i=1,2$ give the number of
2-strings whose imaginary parts $w_k^{(i)}$ are greater than that of the given 1-string $v_j^{(i)}$.
The 1-strings $v_1^{(i)}$ and 2-strings $w_1^{(i)}$ labelled by $1$ are
closest to  the real axis.
The quantum numbers $I^{(i)}_j$ satisfy
\begin{equation}
n_{i}\ge I^{(i)}_{1}\ge I^{(i)}_{2}\ge\dots\ge I^{(i)}_{m_{i}}\ge 
0,\qquad i=1,2.
\label{iranges}
\end{equation}
For given string
content
$(\vm,\vn)$, the lowest excitation occurs when all of the 1-strings 
are further out from
the real axis than all of the 2-strings. In this case  all of the 
quantum numbers vanish
$I_j^{(i)}=0$. Bringing the location of a 1-string closer to the real axis by
interchanging the location of the 1-string with a 2-string increments 
its quantum number by
one unit and increases the energy.

Although we do not make use of it here, we mention in passing that,  for the tricritical Ising model, there exists a bijection~\cite{FP03} between the patterns of 1- and 2-strings classifying the eigenvalues of the double-row transfer eigenvalues and RSOS paths on $A_4$. This has as a consequence that the finitized partition functions (Virasoro characters) satisfy the same recursions~\cite{ABF84} as the one-dimensional configurational sums of the Corner Transfer Matrices (CTMs).

\subsection{Three mechanisms for changing string content}

Let $\ttt(u)$ be the normalized double row transfer matrix as in PCAI. 
In Regime~IV the eigenvalues $t(u)$ of $\ttt(u)$ are doubly periodic meromorphic functions
in the period rectangle
\begin{equation}
\text{period rectangle}=
\big(-\frac{\lambda}{2},\frac{9\lambda}{2}\big)\times
(-\pi i\veps,\pi i\veps)
\end{equation}
where $\veps$ is related to the departure-from-criticality variable $t$ by 
\begin{equation}
t=-\exp(-2\pi\veps).
\end{equation}
However, in Regime~IV the $A_4$ lattice model also admits the symmetry 
\begin{equation}
t(u\pm\pi/2+\pi i\veps)=t(u)\label{xsymm}
\end{equation}
so we can restrict ourselves further to the rectangle
$(-\frac{\lambda}{2},2\lambda)\times (-\pi i\veps,\pi i\veps)$. In particular, this means that
for $mR>0$ the distinction between the two strips effectively disappears --- the two are joined
along the scaling edge at $\Im(u)=\pi\veps/2$ in the scaling limit (\ref{scalinglimit}). This
allows zeros to move between the putative strips~1 and 2 by crossing the scaling edge. The
combination of strips~1 and 2 in the upper half plane into one extended strip is shown in
Figure~1. There is a complex conjugation symmetry between the upper and lower half planes. Note
that this is consistent with the picture at the conformal point since in the limit $mR\to 0$ the
zeros are infinitely far below (strip~1) or infinitely far above (strip~2) the scaling edge as
$\veps\to\infty$.
\begin{figure}[bth]
\vspace{-0.25in}
\begin{center}
\includegraphics[width=.5\linewidth]{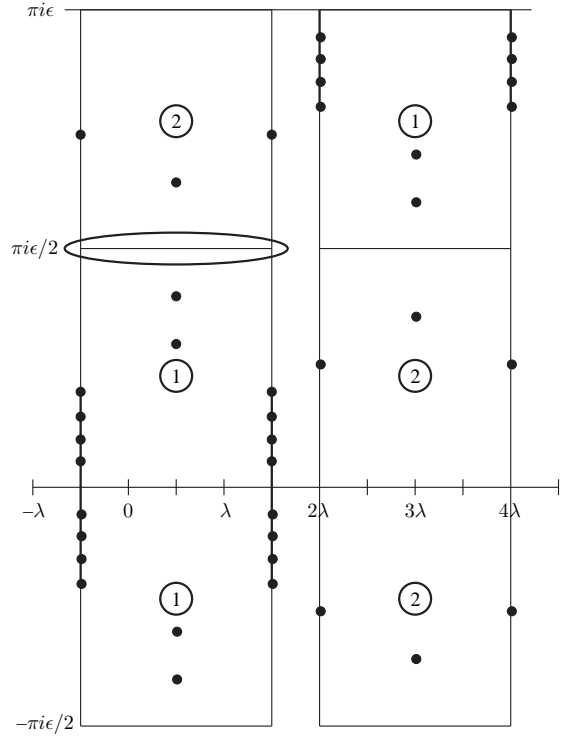}
\end{center}
\vspace{-0.25in}
    \caption{Upper part of the period rectangle 
$(-\frac{\lambda}{2},2\lambda)\times (-\pi i\veps,\pi i\veps)$ in the complex $u$-plane for the
$A_4$ lattice model in Regime~IV showing the putative strips 1 and 2. The schematic location of
zeros reflects the complex conjugation symmetry and the symmetry (\ref{xsymm}).  In the scaling
limit the imaginary period $\veps\to\infty$ and it is the behaviour of the zeros near the indicated scaling edge at
$\Im(u)=\pi\veps/2$ that is relevant.}\label{PeriodRectFig}     
\end{figure}

Let us consider the extended analyticity strip $-\frac{\lambda}{2}\le\Re(u)\le\frac{3\lambda}{2}$.
Empirically, we find that as $mR$ is increased from $0$ the strip~1 zeros approach the scaling
edge at $\Im(u)=\pi\veps/2$ from below while the strip~2 zeros approach the scaling edge from
above. Potentially, this allows zeros to collide or migrate from one strip to the other. In
fact we find that the zero patterns change by one of the following three mechanisms which
occur at the scaling edge:

\newpage
\renewcommand{\theenumi}{\Alph{enumi}}
\begin{enumerate}
\item If 1-strings are closest to the scaling edge in both strips~1 and 2 then these 1-strings
collide and move symmetrically in opposite directions parallel to the scaling edge until they reach $\Re(u)=0,\lambda$. At this point the two zeros no longer contribute and they can be removed from the analyticity strip. This mechanism applies if $I^{(1)}_{m_1}=I^{(2)}_{m_2}=0$.
\item If a 2-string is closest to the scaling edge in strip~2 and a 1-string is closest to the
scaling edge in strip~1 then the 2-string moves to the scaling edge and leaves the analyticity
strip. This mechanism applies if $I^{(1)}_{m_1}=0$ and $I^{(2)}_{m_2}>0$.
\item Otherwise, if a 2-string is closest to the scaling edge in strip~1, then this 2-string
moves to the scaling edge and leaves the analyticity strip. This mechanism applies if
$I^{(1)}_{m_1}> 0$.
\end{enumerate}
Similar mechanisms are also observed~\cite{FPR02} in the boundary flows of the tricritical Ising model. 
If $m_2=0$ only Mechanisms B and C apply according to whether $I^{(1)}_{m_1}=0$ or $I^{(1)}_{m_1}>0$. If $m_1=m_2=0$ then Mechanism C
applies. In each of the three mechanisms exactly two zeros are removed from the extended
analyticity strip. These three mechanisms are shown schematically in Figure~2. They are
confirmed by the numerics discussed in Section~4.

\begin{figure}[htb]
\vspace{-0.75in}
\begin{center}
\includegraphics[width=1.0\linewidth]{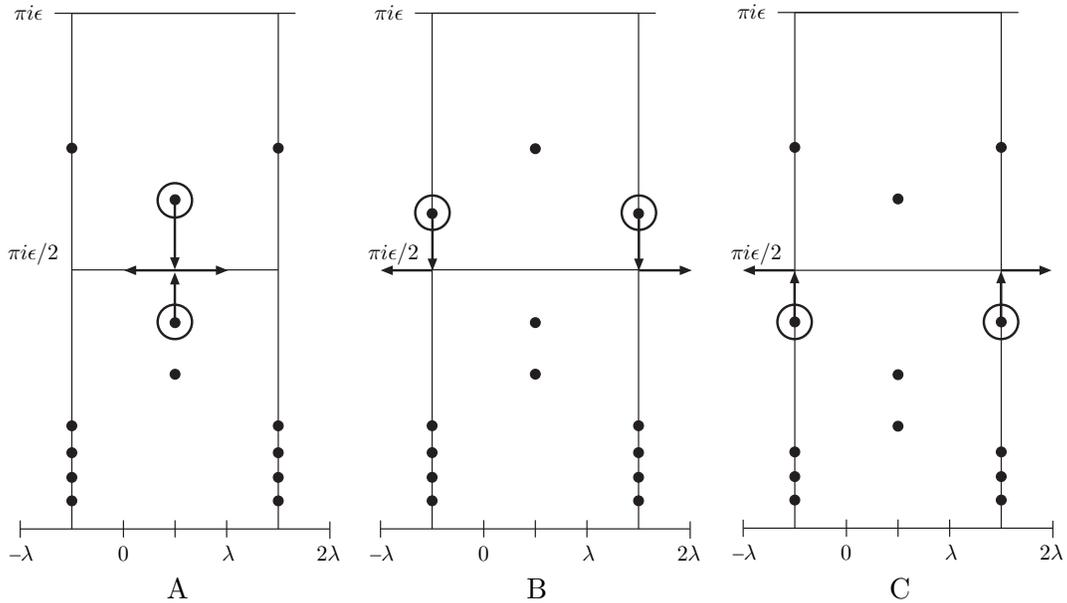}
\end{center}
\vspace{-0.95in}
    \caption{Schematic representation of the three mechanisms A, B, C by which the string
contents change under the flow. In each case the two
circled zeros leave the analyticity strip as indicated by the arrows.
In mechanism A two 1-strings leave the strip whereas in mechanisms B
and C it is a 2-string that leaves the strip. Note that only the
location of the 1-strings enter the TBA equations.}\label{MechanismsFig}     
\end{figure}

\subsection{Operator flow}

Remarkably,  as we explain in this subsection, the empirical rules giving the three
mechanisms A,B,C for  changing string contents under the flow suffice to determine a
map in each sector $(r,s)$ between finitized characters for the UV ($mR=0$, $c=7/10$) and IR
($mR=\infty$, $c=1/2$) fixed points. Since two zeros
leave the analyticity strip under the flow, this map from the UV to IR
takes the form
\begin{equation}
\chi^{4,N+2}_{r,s}(q)\mapsto \chi^{3,N}_{r',s'}(q),\qquad 
1\le r'\le 2,\quad 1\le s',r\le 3,\quad 1\le s\le 4.
\end{equation}
We find that the primary operators $(r,s)$ of the tricritical Ising model flow to the primary
operators $(r',s')$ of the critical Ising model as shown in Table~1. This pattern of flows is
consistent with the flow of operators observed with periodic boundaries but, in the presence
of a boundary, the flow is always to the primary operator not to an associated 
descendant as can occur in the periodic case. 
\begin{table}[htb]
\begin{center}
\begin{tabular}{r|cccl}
\multicolumn{5}{c}{$\Delta^{(4,5)}_{r,s}$}\\
\multicolumn{5}{l}{$s$}\\
4&${3\over 2}$&${7\over 16}$&$0$&\\[4pt]
3&${3\over 5}$&${3\over 80}$&${1\over 10}$&\\[4pt] 
2&${1\over 10}$&${3\over 80}$&${3\over 5}$&\\[4pt] 
1&0&${7\over 16}$&${3\over 2}$&\\[4pt] 
\cline{2-4}
\multicolumn{2}{r}{1} &   2   &   3  &  $r$
\end{tabular}
\quad$\mapsto$\quad
\begin{tabular}{r|cccl}
\multicolumn{5}{c}{$(r',s')$}\\
\multicolumn{5}{l}{$s$}\\
4&(1,3)&(1,2)&(1,1)&\\[4pt] 
3&(2,3)&(2,2)&(2,1)&\\[4pt] 
2&(2,1)&(2,2)&(2,3)&\\[4pt] 
1&(1,1)&(1,2)&(1,3)&\\[4pt] 
\cline{2-4}
\multicolumn{2}{r}{1\phantom{$1)$}} &   2   &   3  &  $r$
\end{tabular}
$=$\quad
\begin{tabular}{r|cccl}
\multicolumn{5}{c}{$\Delta^3_{r',s'}$}\\
\multicolumn{5}{l}{$s$}\\
4&${1\over 2}$&${1\over 16}$&$0$&\\[4pt] 
3&$0$&${1\over 16}$&${1\over 2}$&\\[4pt] 
2&${1\over 2}$&${1\over 16}$&$0$&\\[4pt] 
1&0&${1\over 16}$&${1\over 2}$&\\[4pt] 
\cline{2-4}
\multicolumn{2}{r}{1}   &   2   &   3  &  $r$
\end{tabular}
\end{center}
\caption{Flow of primary operators $(r,s)\mapsto (r',s')$ displayed in the $A_4$ Kac table.}
\end{table}

\subsection{RG flow from $\chi^4_{1,1}(q)$ to $\chi^3_{1,1}(q)$}

In the vacuum sector $(r,s)=(1,1)$ the $A_4$ $(\vm,\vn)_{N+2}$ system of the tricritical Ising model is
\begin{equation}
m_1+n_1=\half(N+2+m_2),\qquad m_2+n_2=\half m_1
\end{equation} 
and these relations determine $n_1$ and $n_2$ in terms of $m_1$ and $m_2$. 
Similarly, let $m$ and $n$ be the number of 1- and 2-strings of the critical
Ising model in the vacuum $(r,s)=(1,1)$ sector satisfying the $A_3$ $(\vm,\vn)_N$ system 
\begin{equation}
m+n=\half(N+m)\quad \text{or}\quad n=\half(N-m).
\end{equation}
Then $m$ and $n$ are given by the total number of 1-strings and 2-strings in the extended strip
in the IR limit
\begin{equation}
m=
\begin{cases}
m_1+m_2-2,&\mbox{A}\\ 
m_1+m_2,&\mbox{B,C}
\end{cases}
\qquad\qquad n=
\begin{cases}
n_1+n_2,&\mbox{A}\\
n_1+n_2-1,&\mbox{B,C}.
\end{cases}
\label{Isingmn}
\end{equation}

\begin{table}[htbp]
\begin{center}
\begin{tabular}{|c|c|c|c|c|c||c|c|c|}\hline\hline
\rule[-.2cm]{0cm}{.65cm}Mech&$E$&$\ m_1\ $&$\ m_2\ $&
$n_2$&$I$&$m$&$I'$&$E'$\\ \hline\hline
\rule[-.2cm]{0cm}{.65cm}C&0&0&0&0&$()$&0&$()$&0\\ \hline
\rule[-.2cm]{0cm}{.65cm}B&2&2&0&1&$(0,0)$&2&$(0,0)$&2\\ \hline
\rule[-.2cm]{0cm}{.65cm}B&3&2&0&1&$(1,0)$&2&$(1,0)$&3\\ \hline
\rule[-.2cm]{0cm}{.65cm}B&4&2&0&1&$(2,0)$&2&$(2,0)$&4\\ \hline
\rule[-.2cm]{0cm}{.65cm}B&4&2&0&1&$(1,1)$&2&$(1,1)$&4\\ \hline
\rule[-.2cm]{0cm}{.65cm}B&5&2&0&1&$(3,0)$&2&$(3,0)$&5\\ \hline
\rule[-.2cm]{0cm}{.65cm}B&5&2&0&1&$(2,1)$&2&$(2,1)$&5\\ \hline
\rule[-.2cm]{0cm}{.65cm}B&6&2&0&1&$(4,0)$&2&$(4,0)$&6\\ \hline
\rule[-.2cm]{0cm}{.65cm}B&6&2&0&1&$(3,1)$&2&$(3,1)$&6\\ \hline
\rule[-.2cm]{0cm}{.65cm}B&6&2&0&1&$(2,2)$&2&$(2,2)$&6\\ \hline
\rule[-.2cm]{0cm}{.65cm}A&6&4&2&0&$(0,0,0,0|0,0)$&4&$(0,0,0,0)$&8\\ \hline
\rule[-.2cm]{0cm}{.65cm}B&7&2&0&1&$(5,0)$&2&$(5,0)$&7\\ \hline
\rule[-.2cm]{0cm}{.65cm}B&7&2&0&1&$(4,1)$&2&$(4,1)$&7\\ \hline
\rule[-.2cm]{0cm}{.65cm}B&7&2&0&1&$(3,2)$&2&$(3,2)$&7\\ \hline
\rule[-.2cm]{0cm}{.65cm}A&7&4&2&0&$(1,0,0,0|0,0)$&4&$(1,0,0,0)$&9\\ \hline
\rule[-.2cm]{0cm}{.65cm}B&8&2&0&1&$(6,0)$&2&$(6,0)$&7\\ \hline
\rule[-.2cm]{0cm}{.65cm}B&8&2&0&1&$(5,1)$&2&$(5,1)$&7\\ \hline
\rule[-.2cm]{0cm}{.65cm}B&8&2&0&1&$(4,2)$&2&$(4,2)$&7\\ \hline
\rule[-.2cm]{0cm}{.65cm}B&8&2&0&1&$(3,3)$&2&$(3,3)$&7\\ \hline
\rule[-.2cm]{0cm}{.65cm}A&8&4&2&0&$(2,0,0,0|0,0)$&4&$(2,0,0,0)$&10\\ \hline
\rule[-.2cm]{0cm}{.65cm}A&8&4&2&0&$(1,1,0,0|0,0)$&4&$(1,1,0,0)$&10\\ \hline
\rule[-.2cm]{0cm}{.65cm}B&8&4&0&2&$(0,0,0,0)$&4&$(1,1,1,1)$&12\\ \hline
\end{tabular}
\end{center}
\caption{Mapping of the first 22 energy levels in the $(r,s)=(1,1)$ sector under Mechanisms A, B or C.
The energies, string contents and quantum numbers are shown in both the UV and IR regimes.
The degeneracies of all the levels reconstruct the mapping between the tricritical and critical Ising
characters $\chi_{1,1}^4(q)\mapsto \chi_{1,1}^3(q)$.} 
\end{table}

The mechanisms A, B, C determine which energy level flows to which energy level under the RG
flow. As explained in \cite{OPW97} and PCAI, the energies at a conformal point are 
determined by the string content $\vm$ and the patterns of zeros in the complex $u$-plane.
In terms of quantum numbers the precise mapping of energy levels under the flow is given by
\begin{eqnarray}
I=(I_1^{(1)},I_2^{(1)},\ldots,I_{m_1}^{(1)}\;|\;I_1^{(2)},I_2^{(2)},
\ldots,I_{m_2}^{(2)})\mapsto I'=(I_1',I_2',\ldots,I_m')
\label{Isingquantum}
\end{eqnarray}
where
\begin{eqnarray}
\mbox{A:\phantom{,C}}\qquad &&
\begin{cases}
I_j'=n_2+I^{(1)}_j,&\quad j=1,2,\ldots,m_1-1\\
I_{m_1-1+k}'=n_2-I^{(2)}_{m_2-k},&\quad k=1,2,\ldots,m_2-1
\end{cases}\label{Isingquantum1}\\
\mbox{B,C:}\qquad&&
\begin{cases}
I_j'=n_2-1+I^{(1)}_j,&\qquad j=1,2,\ldots,m_1\\
I_{m_1+k}'=n_2-I^{(2)}_{m_2+1-k},&\qquad k=1,2,\ldots,m_2
\end{cases}
\label{Isingquantum2}
\end{eqnarray}
Details of this mapping for the first 15 energy levels in the $(r,s)=(1,1)$ sector are shown in Table~2.
For given string content $\vm$, the base energy level $E_m$ is
determined by the Cartan matrix
$C$
\begin{equation}
q^{E_m}=q^{{1\over 4}\vm C\vm}=
\begin{cases}
q^{{1\over 2}(m_1^2-m_1m_2+m_2^2)},&A_4\\
q^{{1\over 2} m^2},&A_3.
\end{cases}
\end{equation}
The base energy occurs when the location of all of the 1-strings are further from the real axis than the
locations of all of the 2-strings.  Additional excitation energy is generated by permuting the
order of 1-strings and 2-strings in each strip as dictated by the sum of quantum numbers $I_j$ in
each strip and is given by the Gaussian polynomials or $q$-binomials
\begin{equation}
\gauss{m+n}{m}=\gauss{m+n}{m}_q=
\sum_{I_1=0}^n \sum_{I_2=0}^{I_1}\cdots \sum_{I_m=0}^{I_{m-1}} q^{I_1+\ldots+I_m}
=\begin{cases}
\dfrac{(q)_{m+n}}{(q)_{m}(q)_{n}}, &\quad m,n\ge 0 \\
0, &\quad \text{otherwise}\\
\end{cases}
\end{equation}
with the $q$-factorials $(q)_{m}=(1-q)\cdots(1-q^{m})$ for $m\ge 1$ 
and $(q)_{0}=1$. 
The energy $E$ is increased by one unit each time a 1-string is brought closer to the real axis
by interchanging its location with the location of an adjacent 2-string. The product of
two $q$-binomials is the generating function for the conformal spectra with given string content
in each strip. The
$q$-binomials satisfy the properties
\begin{equation}
\gauss{n}{m}=\gauss{n}{n-m}=\gauss{n-1}{m-1}+q^m\gauss{n-1}{m},\qquad
\gauss{m+n}{m}_q=q^{mn}\gauss{m+n}{m}_{1\over q}.\label{qbinprops}
\end{equation}

The character $\chi^4_{1,1}(q)$ is the generating function for the tricritical Ising conformal
spectra in the $(1,1)$ sector. Explicitly, using the recursion (\ref{qbinprops}), we decompose it
into three terms 
\begin{eqnarray}
\chi_{1,1}^{4,N+2}(q)\!\!\!
&=&\!\!\!\!q^{-{7\over 240}}\sum_{(\vm,\vn)_{N+2}} q^{{1\over 4}\vm C\vm}
\gauss{m_1+n_1}{m_1}\gauss{m_2+n_2}{m_2}\nonumber\\
&=&\!\!\!\!q^{-{7\over 240}}
\sum_{\mbox{\tiny $m_1$,$m_2$ even}}\!\! q^{{1\over 2}(m_1^2-m_1m_2+m_2^2)}
\left\{\gauss{m_1+n_1-1}{m_1-1}\gauss{m_2+n_2-1}{m_2-1}\right.\\
&&\qquad\mbox{}+\left.
q^{m_2}\gauss{m_1+n_1-1}{m_1-1}\gauss{m_2+n_2-1}{m_2}+
q^{m_1}\gauss{m_1+n_1-1}{m_1}\gauss{m_2+n_2}{m_2}\nonumber
\right\}
\end{eqnarray}
These three terms correspond precisely to the energy levels effected by
mechanisms A, B and C respectively. 
So simply reading off the conformal energies from the
respective zero patterns after applying each mechanism we find the following mapping
between finitized characters
\begin{eqnarray}
\chi_{1,1}^{4,N+2}(q)\!\!\!
&\mapsto&\!\!q^{-{1\over 48}}\!\!\!\!
\sum_{\mbox{\tiny $m_1$,$m_2$ even}}\!\!\!\!  q^{{1\over 2}m^2} \!
\left\{ q^{n_2(m_1-1)+n_2(m_2-1)}
\gauss{m_1+n_1-1}{m_1-1}\gauss{m_2+n_2-1}{m_2-1}_{1\over q}
\right.\nonumber\\
&&\mbox{}\hspace{-.7in}\mbox{}+\left.
q^{(n_2-1)m_1+(n_2-1)m_2}\gauss{m_1+n_1-1}{m_1-1}\gauss{m_2+n_2-1}{m_2}_{1\over q}+
q^{n_2m_1+n_2m_2}\gauss{m_1+n_1-1}{m_1}\gauss{m_2+n_2}{m_2}_{1\over q}
\right\}\nonumber\\
&=&\!\!\!q^{-{1\over 48}}\!\sum_{\mbox{\tiny $m$ even}} q^{{1\over 2}m^2} \!\!
\sum_{\mbox{\tiny $m_1$ even}}
\left\{ q^{n_2(m_1-1)}
\gauss{m_1+n_1-1}{m_1-1}\gauss{m_2+n_2-1}{m_2-1}
\right.\\
&&\qquad\mbox{}+\left.
q^{(n_2-1)m_1}\gauss{m_1+n_1-1}{m_1-1}\gauss{m_2+n_2-1}{m_2}+
q^{n_2m_1}\gauss{m_1+n_1-1}{m_1}\gauss{m_2+n_2}{m_2}
\right\}\nonumber\\
&=&\!\!\!\!q^{-{1\over 48}}\!\sum_{\mbox{\tiny $m$ even}} q^{{1\over 2}m^2} \!\!
\sum_{\mbox{\tiny $m_1$ even}} q^{{3\over 2}m_1^2-mm_1}
\left\{q^{m-{7\over 2}m_1+2}\gauss{{N+m-m_1+2\over 2}}{m_1-1}\gauss{{m_1-2\over 2}}{m-m_1+1}
\right.\nonumber\\
&&\qquad\qquad\qquad\qquad\left.\mbox{}+
q^{-m_1}\gauss{{N+m-m_1\over 2}}{m_1-1}\gauss{{m_1-2\over 2}}{m-m_1}+
\gauss{{N+m-m_1\over 2}}{m_1}\gauss{{m_1\over 2}}{m-m_1}\right\}\nonumber\\
&=&\!\!\!\!q^{-{1\over 48}}\!
\sum_{\mbox{\tiny $m$ even}} q^{{1\over 2}m^2}\gauss{{N+m\over 2}}{m}
\;=\;\chi_{1,1}^{3,N}(q)\nonumber
\end{eqnarray}
Notice that, after the mapping, the $q$-binomials of strip~2 are with respect to $1/q$. This is 
because strip~2 is turned upside down when it is placed on top of strip~1 to form the extended
strip. These
$q$-binomials are naturally replaced by $q$-binomials in $q$ using (\ref{qbinprops}). All
integers are then eliminated in favour of $m$ and $m_1$ using the appropriate relations  which
apply to the mechanism corresponding to each of the three terms. The final equality holds
because of the remarkable generalized $q$-Vandermonde identity which is proved in the Appendix
\begin{eqnarray}
\gauss{{N+m\over 2}}{m}&=&\sum_{\mbox{\tiny $m_1$ even}} q^{{3\over 2}m_1^2-mm_1}
\left\{q^{m-{7\over 2}m_1+2}\gauss{{N+m-m_1+2\over 2}}{m_1-1}\gauss{{m_1-2\over 2}}{m-m_1+1}
\right.\\
&&\qquad\qquad\left.\mbox{}+
q^{-m_1}\gauss{{N+m-m_1\over 2}}{m_1-1}\gauss{{m_1-2\over 2}}{m-m_1}+
\gauss{{N+m-m_1\over 2}}{m_1}\gauss{{m_1\over 2}}{m-m_1}\right\}.\nonumber
\end{eqnarray}

The finitized Ising character $\chi_{1,1}^{3,N}(q)$ is not the usual finitized character. In the
limit $q\to 1$ it gives the correct $A_4$ counting
\begin{equation}
\lim_{q\to 1} \chi_{1,1}^{3,N}(q)
=\sum_{\mbox{\tiny $m$ even}}\mbox{\small ${{N+m\over 2}\choose m}$}
=\big[A_4^{N+2}\big]_{1,1}
\end{equation}
whereas the usual finitized Ising
character gives $A_3$ counting
\begin{equation}
\lim_{q\to 1} q^{-{1\over 48}}\!
\sum_{\mbox{\tiny $m$ even}} q^{{1\over 2}m^2}\gauss{{N\over 2}}{m}
=\sum_{\mbox{\tiny $m$ even}}\mbox{\small ${{N\over 2}\choose m}$}
=\big[A_3^N\big]_{1,1}
\end{equation}
where $A_4$ and $A_3$ denote the adjacency matrices.

\newpage
\subsection{RG flow in other sectors}

The analysis of the flow using the three mechanisms A, B and C can be extended to each of the
sectors $(r,s)$ with $r=1,2$, $s=1,2,3$. In this way we obtain six generalized 
\mbox{$q$-Vandermonde} identities as follows:
\begin{eqnarray}
&&\chi_{1,1}\qquad N\ \mbox{even}, m\ \mbox{even:}\nonumber\\
&&\quad\qquad\gauss{(N+m)/2}{m}=\sum_{m_1\,\text{even}} q^{3m_1^2/2-m m_1}\left\{
               \gauss{(N+m-m_1)/2}{m_1}     \gauss{m_1/2}{m-m_1}\right.\\ 
&&\left.\mbox{}+
q^{-m_1}        \gauss{(N+m-m_1)/2}{m_1-1}   \gauss{(m_1-2)/2}{m-m_1}+
q^{m-7m_1/2+2}		\gauss{(N+m-m_1+2)/2}{m_1-1} \gauss{(m_1-2)/2}{m-m_1+1}\right\}\nonumber\\[8pt]
&&\chi_{3,1}\qquad N\ \mbox{even}, m\ \mbox{odd:}\nonumber\\
&&\quad\qquad\gauss{(N+m+1)/2}{m}=\sum_{m_1\,\text{even}} q^{3m_1^2/2-m m_1}\left\{
               \gauss{(N+m-m_1+1)/2}{m_1}   \gauss{m_1/2}{m-m_1}\right.\\ 
&&\left.\mbox{}+
q^{-m_1}        \gauss{(N+m-m_1+1)/2}{m_1-1} \gauss{(m_1-2)/2}{m-m_1}+
q^{m-7m_1/2+2}		\gauss{(N+m-m_1+3)/2}{m_1-1} \gauss{(m_1-2)/2}{m-m_1+1}\right\}\nonumber\qquad\qquad\qquad
\end{eqnarray}
\begin{eqnarray}
&&\chi_{2,2}\qquad N\ \mbox{even}, m\ \mbox{even:}\nonumber\\
&&\quad\qquad\gauss{(N+m)/2}{m}=\sum_{m_1\,\text{even}} q^{3m_1^2/2-m m_1}\left\{
				           \gauss{(N+m-m_1)/2}{m_1}   \gauss{m_1/2}{m-m_1}\right.\\ 
&&\left.\mbox{}+
q^{-m+5m_1/2+1} \gauss{(N+m-m_1-2)/2}{m_1} \gauss{m_1/2}{m-m_1-1}+
q^{m-5m_1/2+1}  \gauss{(N+m-m_1)/2}{m_1-1} \gauss{m_1/2}{m-m_1+1}\right\}\nonumber\qquad\qquad\\[8pt]
&&\chi_{2,1}\qquad N\ \mbox{odd}, m\ \mbox{even:}\nonumber\\
&&\quad\qquad\gauss{(N+m-1)/2}{m}=\sum_{m_1\,\text{odd}} q^{3m_1^2/2-m m_1}\left\{
q^{m_1/2}		      \gauss{(N+m-m_1-2)/2}{m_1}    \gauss{(m_1+1)/2}{m-m_1}\right.\\ 
&&\left.\mbox{}+
q^{-m_1/2}       \gauss{(N+m-m_1-2)/2}{m_1-1}  \gauss{(m_1-1)/2}{m-m_1}+
q^{m-3m_1+3/2}   \gauss{(N+m-m_1)/2}{m_1-1}    \gauss{(m_1-1)/2}{m-m_1+1}\right\}\nonumber
\end{eqnarray}
\begin{eqnarray}
&&\chi_{1,2}\qquad N\ \mbox{odd}, m\ \mbox{even:}\nonumber\\
&&\quad\qquad\gauss{(N+m-1)/2}{m}=\sum_{m_1\,\text{odd}} q^{3m_1^2/2-m m_1}\left\{
q^{m-3m_1+3/2}		 \gauss{(N+m-m_1)/2}{m_1-1}  \gauss{(m_1-1)/2}{m-m_1+1}\right.\\ 
&&\left.\mbox{}+
q^{-m+2m_1+1/2}  \gauss{(N+m-m_1-2)/2}{m_1}  \gauss{(m_1-1)/2}{m-m_1-1}+
q^{-m_1/2}       \gauss{(N+m-m_1)/2}{m_1}    \gauss{(m_1-1)/2}{m-m_1}\right\}\nonumber\\[8pt]
&&\chi_{3,2}\qquad N\ \mbox{odd}, m\ \mbox{odd:}\nonumber\\
&&\quad\qquad\gauss{(N+m)/2}{m}=\sum_{m_1\,\text{odd}} q^{3m_1^2/2-m m_1}\left\{
q^{m-3m_1+3/2}		 \gauss{(N+m-m_1+1)/2}{m_1-1}  \gauss{(m_1-1)/2}{m-m_1+1}\right.\\ 
&&\left.\mbox{}+
q^{-m+2m_1+1/2}  \gauss{(N+m-m_1-1)/2}{m_1}    \gauss{(m_1-1)/2}{m-m_1-1}+
q^{-m_1/2}       \gauss{(N+m-m_1+1)/2}{m_1}    \gauss{(m_1-1)/2}{m-m_1}\right\}\nonumber\qquad\qquad\qquad
\end{eqnarray}
These identities are simplified and proved in the Appendix.

\section{TBA Equations in Regime~IV}

Recall from PCAI that the normalized double row transfer matrix is defined by
\begin{equation}
\label{e:t}
\ttt(u) = S_{r,s}(u)S(u)
\Bigl[i\,\frac{\h_1(u+2\l,p)\h_1(\l,p)}{\h_1(u+3\l,p)\h_1(u+\l,p)}\Bigr]^{2N}
\D(u)
\end{equation}
where
\begin{equation}
S(u)=\frac{\h_1(2u-\l,p)^2}{\h_1(2u-3\l,p)\h_1(2u+\l,p)}
\end{equation}
and
\begin{equation}
S_{r,s}(u) 
  = (-1)^{s}h_r(u-\xL)h_{-r}(u+\xL)\bar{h}_{s}(u-\xR)\bar{h}_{-s}(u+\xR)
\end{equation}
with
\begin{align}
h_r(u)& = \frac{\h_1(\l,p)\h_1(u+(3-r)\l,p)\h_1(u+(1-r)\l,p)}
{\h_1(u,p)\h_1(u-\l,p)\h_1(u+2\l,p)}\\
\bar{h}_s(u)& = \frac{\h_4(\l,p)\h_4(u+(3-s)\l,p)\h_4(u+(1-s)\l,p)}
{\h_4(u,p)\h_4(u-\l,p)\h_4(u+2\l,p)}.
\end{align}
Moreover, the normalized transfer matrix 
satisfies~\cite{BPO96} the
universal TBA functional equation
\begin{equation}
\label{e:functional}
\ttt(u)\ttt(u+\l) = \I + \ttt(u+3\l)
\end{equation}
independent of the boundary condition $(r,s)$. For Regime~IV, the nome is pure imaginary
with $p=ie^{-\pi\u}$.

\subsection{UV Massless TBA: $(r,s)=(1,1)$}

In this section we derive the TBA equations for the $(r,s)=(1,1)$ boundary by
solving the TBA functional equations in the scaling limit for even 
$N$. We follow
closely the derivations in \cite{OPW97,PearN98} and PCAI~\cite{PCAI}. 
The derivation for other boundary conditions is similar. 
We begin by
factorizing the eigenvalue $t(u)$ of $\ttt(u)$ for large $N$ as
\begin{equation}
t(u) = f(u)g(u)l(u)
\end{equation}
where $f(u)$ accounts for the bulk order-$N$ behaviour, $g(u)$ the
order-$1$ boundary contributions and $l(u)$ is the order-$1/N$
finite-size correction. We will solve for $f(u)$, $g(u)$ and then $l(u)$
sequentially. 

For the order-$N$ behaviour the second term on the RHS of the TBA
functional equation \eqref{e:functional} can be neglected giving the inversion
relation
\begin{equation}\label{e:bulkf}
f(u)f(u+\l) = 1.
\end{equation}
In the physical strip~1, the solution~\cite{Bax82Inv} with the required analyticity is
\begin{equation}
f(u) = \Bigl[\frac{\h_4(\frac{\pi}{4}-\frac{5u}{2},|t|^{5/2})}
{\h_4(\frac{\pi}{4}+\frac{5u}{2},|t|^{5/2})}\Bigr]^{2N}.
\end{equation}
The solution in strip~2 satisfies the same inversion relation and is given  by the symmetry \eqref{xsymm}.

In the two analyticity strips labelled by $j=1,2$ we define generically for the
functions $h=t,f,g,l$ the notations
\begin{subequations}
\begin{align}
h_1(x)& = h(\frac{\l}{2}+\frac{ix}{5}), \quad \;\text{$|\Im(x)|<\pi$}\\
h_2(x)& = h(3\l+\frac{ix}{5}), \quad \text{$|\Im(x)|<\pi$}\\[5pt]
H_1(x) =& 1+h_1(x), \quad H_2(x) = 1+h_2(x).
\end{align}
\end{subequations}
and we assume the relevant functions have the scaling form
\begin{equation}
\hat{h}_i(x)=\lim_{N\to\infty} h(x+\log N)
\end{equation}
For example, we see that
\begin{equation}
\log \hat{f}_1(x)=-4\mu^2e^x
\end{equation}

As in Regime~III, we next have to solve by Fourier series the inversion relations for the
order-1 boundary terms
\begin{subequations}
\begin{align}\label{e:g3}
g_1(x-i\frac{\pi}{2})g_1(x+i\frac{\pi}{2}) & = 1\\
g_2(x-i\frac{\pi}{2})g_2(x+i\frac{\pi}{2}) &= G_1(x).
\end{align}
\end{subequations}
To find $g_i(x)$ we need to consider the zeros and poles
introduced by the prefactor in \eqref{e:t}.
We find the order-1 zeros of $D(u)$ cancel exactly the poles of $S_{1,1}(u)$.
Taking into account periodicity, the prefactor exhibits poles at $u=-{\l\over 2}+\sigma i\pi\u$,
${3\l\over 2}+\sigma i\pi\u$, $2\l+\sigma i\pi\u$, $4\l+\sigma i\pi\u$,
${\l\over 4}\pm i{\pi\u\over 2}$, ${3\l\over 4}\pm i{\pi\u\over 2}$, 
${11\l\over 4}\pm i{\pi\u\over 2}$, ${13\l\over 4}\pm i{\pi\u\over 2}$ and double zeros at
$u={\l\over 2}+ i\sigma\pi\u$, $3\l+i\sigma \pi\u$, ${7\l\over 4}\pm i{\pi\u\over 2}$, 
${17\l\over 4}\pm i{\pi\u\over 2}$ where
$\sigma=0,\pm 1$. The solution for $g_1(x)$ with this analyticity but restricted to
the strip $|\Im(x)|<3\pi/4$ is
\begin{equation}\label{e:g11}
g_1(x) = -\frac{
\h_3({ix\over 2}-{\pi\over 8},|t|^{5/4})
\h_3({ix\over 2}+{\pi\over 8},|t|^{5/4})}
{\h_4({ix\over 2}-{\pi\over 8},|t|^{5/4})
\h_4({ix\over 2}+{\pi\over 8},|t|^{5/4})}
\Bigl[\frac{\h_1({ix\over 2},|t|^{5/4})}
{\h_2({ix\over 2},|t|^{5/4})}\Bigr]^2.
\end{equation}

By comparing the expected pattern of zeros and poles of
$g_2(x-i\pi/2)g_2(x+i\pi/2)$
with the analyticity of $G_1(x)$, we find that this functional
relation is satisfied inside the strip $|\Im(x)|<{3\pi\over 4}$. We 
observe that $g_2(x)/g_1(x)$ is free of zeros and poles
in this strip.
Similarly $G_1(x)$ is analytic and nonzero, but only inside
a narrower strip $|\Im(x)|<{\pi\over 4}$.
Hence in this smaller strip 
\begin{equation}
\log{g_2(x)} = \log{g_1(x)} + \veps*\log{G_1}(x)
\end{equation}
where the kernel in Regime~IV is given by
\begin{equation}
\veps(x) = \frac{\h_2(0,|t|^{5})\h_3(0,|t|^{5})\h_3(ix,|t|^{5})}
{2\pi\h_2(ix,|t|^{5})}.
\label{kernel}
\end{equation}
Note that the signs of $g_i(x)$ are choosen to match the corresponding
expressions~\cite{OPW97}
in the tricritical limit $t\to 0$. They can be determined numerically from the eigenvalues of
the transfer matrix with real $u$ in the physical range $0<u<\l/2$. 

The functional equations for the finite-size corrections
\begin{subequations}\label{e:l3}
\begin{align}
l_1(x-\frac{i\pi}{2})l_1(x+\frac{i\pi}{2}) &= T_2(x)\\
l_2(x-\frac{i\pi}{2})l_2(x+\frac{i\pi}{2}) &= \frac{T_1(x)}{G_1(x)}
\end{align}
\end{subequations}
can be converted to Nonlinear Integral Equations (NLIE) by standard techniques~\cite{KlumP91,KlumP92} where the key input is the analyticity determined by the patterns of zeros. 
Suppose that 1-strings are located in the upper half plane in 
the extended strip~1 
at $\{\l/2+iv_1,...,\l/2+iv_{m_1}\}$ below the scaling edge and 
at $\{\l/2+iw_1,...,\l/2+iw_{m_2}\}$ above the scaling edge. Then the zeros in strip~2 are determined by the symmetry \eqref{xsymm} and occur at 
$\{3\l+i(\pi\varepsilon-v_1),...,3\l+i(\pi\varepsilon-v_{m_1})\}$ and 
$\{3\l+i(\pi\varepsilon-w_1),...,3\l+i(\pi\varepsilon-w_{m_2})\}$.
To account for these zeros, we note that the functions
\begin{eqnarray}
l_1(x)\prod_{j=1}^{m_1}\frac{
\h_2(i\frac{x}{2}-i\frac{5v_j}{2},|t|^{\frac{5}{2}})
\h_2(i\frac{x}{2}+i\frac{5v_j}{2},|t|^{\frac{5}{2}})}
{\h_1(i\frac{x}{2}-i\frac{5v_j}{2},|t|^{\frac{5}{2}})
\h_1(i\frac{x}{2}+i\frac{5v_j}{2},|t|^{\frac{5}{2}})}
\prod_{k=1}^{m_2}\frac{
\h_2(i\frac{x}{2}-i\frac{5w_k}{2},|t|^{\frac{5}{2}})
\h_2(i\frac{x}{2}+i\frac{5w_k}{2},|t|^{\frac{5}{2}})}
{\h_1(i\frac{x}{2}-i\frac{5w_k}{2},|t|^{\frac{5}{2}})
\h_1(i\frac{x}{2}+i\frac{5w_k}{2},|t|^{\frac{5}{2}})}&&\ \\
l_2(x)\prod_{j=1}^{m_1}\frac{
\h_3(i\frac{x}{2}-i\frac{5v_j}{2},|t|^{\frac{5}{2}})
\h_3(i\frac{x}{2}+i\frac{5v_j}{2},|t|^{\frac{5}{2}})}
{\h_4(i\frac{x}{2}-i\frac{5v_j}{2},|t|^{\frac{5}{2}})
\h_4(i\frac{x}{2}+i\frac{5v_j}{2},|t|^{\frac{5}{2}})}
\prod_{k=1}^{m_2}\frac{
\h_3(i\frac{x}{2}-i\frac{5w_k}{2},|t|^{\frac{5}{2}})
\h_3(i\frac{x}{2}+i\frac{5w_k}{2},|t|^{\frac{5}{2}})}
{\h_4(i\frac{x}{2}-i\frac{5w_k}{2},|t|^{\frac{5}{2}})
\h_4(i\frac{x}{2}+i\frac{5w_k}{2},|t|^{\frac{5}{2}})}&&\ 
\end{eqnarray}
are free of zeros and poles inside their respective analyticity strips. The products of elliptic functions satisfy the inversion relations $l_i(x-\frac{i\pi}{2})l_i(x+\frac{i\pi}{2}) =1$, $i=1,2$ and are doubly-periodic.

 It follows that
\begin{eqnarray}\label{e:logt4}
\log{t_1(x)}&=& \log{f_1(x)} + \log{g_1(x)} + {\veps}*\log{T_2}(x)+C_1\\
&&\hspace{-1.0in}\mbox{}+ \sum_{j=1}^{m_1}\log
\frac{
\h_1(i\frac{x}{2}-i\frac{5v_j}{2},|t|^{\frac{5}{2}})
\h_1(i\frac{x}{2}+i\frac{5v_j}{2},|t|^{\frac{5}{2}})}
{\h_2(i\frac{x}{2}-i\frac{5v_j}{2},|t|^{\frac{5}{2}})
\h_2(i\frac{x}{2}+i\frac{5v_j}{2},|t|^{\frac{5}{2}})}
+ \sum_{k=1}^{m_2}\log\frac{
\h_1(i\frac{x}{2}-i\frac{5w_k}{2},|t|^{\frac{5}{2}})
\h_1(i\frac{x}{2}+i\frac{5w_k}{2},|t|^{\frac{5}{2}})}
{\h_2(i\frac{x}{2}-i\frac{5w_k}{2},|t|^{\frac{5}{2}})
\h_2(i\frac{x}{2}+i\frac{5w_k}{2},|t|^{\frac{5}{2}})}
\nonumber\\[10pt]
\log{t_2(x)}&= &\log{f_2(x)} + \log{g_2(x)} + {\veps}*\log{T_1}(x)
-{\veps}*\log{G_1}(x)+C_2\\
&&\hspace{-1.0in}\mbox{}+ \sum_{j=1}^{m_1}\log\frac{
\h_4(i\frac{x}{2}-i\frac{5v_j}{2},|t|^{\frac{5}{2}})
\h_4(i\frac{x}{2}+i\frac{5v_j}{2},|t|^{\frac{5}{2}})}
{\h_3(i\frac{x}{2}-i\frac{5v_j}{2},|t|^{\frac{5}{2}})
\h_3(i\frac{x}{2}+i\frac{5v_j}{2},|t|^{\frac{5}{2}})}
+ \sum_{k=1}^{m_2}\log\frac{
\h_4(i\frac{x}{2}-i\frac{5w_k}{2},|t|^{\frac{5}{2}})
\h_4(i\frac{x}{2}+i\frac{5w_k}{2},|t|^{\frac{5}{2}})}
{\h_3(i\frac{x}{2}-i\frac{5w_k}{2},|t|^{\frac{5}{2}})
\h_3(i\frac{x}{2}+i\frac{5w_k}{2},|t|^{\frac{5}{2}})}
\nonumber
\end{eqnarray}
where $\veps(x)$ is the kernel \eqref{kernel}. 
As in \cite{OPW97}, the integration constants $C_i$ vanish.
So now taking the scaling limit yields
\begin{eqnarray}\label{e:logt5}
\log{\hat{t}_1(x)} &=& -4{\mu}^2e^{x}
+\log{\frac{\mu{e^x}+\sqrt{2}+\mu^{-1}e^{-x}}
{\mu{e^x}-\sqrt{2}+\mu^{-1}e^{-x}}}+{k}*\log{\hat{T}_2}(x)\nonumber\\
&&\hspace{-1.0in}\mbox{}+\sum_{j=1}^{m_1} \log\tanh\Bigl[
\frac{1}{2}(\b^{(1)}_j-x-\log\mu)\Bigr] 
+ \sum_{k=1}^{m_2}\log\tanh\Bigl[
\frac{1}{2}(\b^{(2)}_k-x-\log\mu)\Bigr]\\
\log{\hat{t}_2(x)} &=& -4e^{-x}
+\log{\frac{\mu{e^x}+\sqrt{2}+\mu^{-1}e^{-x}}
{\mu{e^x}-\sqrt{2}+\mu^{-1}e^{-x}}} + {k}*\log{\hat{T}_1}(x)
\nonumber\\
&&\hspace{-1.0in}\mbox{}+\sum_{j=1}^{m_1} \log\tanh\Bigl[
\frac{1}{2}(\b^{(1)}_j+x+\log\mu)\Bigr]
+ \sum_{k=1}^{m_2}\log\tanh\Bigl[
\frac{1}{2}(\b^{(2)}_j+x+\log\mu)\Bigr]
\end{eqnarray}
where the kernel is
\begin{equation}
k(x)=\frac{1}{2\pi\cosh{x}}.
\end{equation}
In deriving this result we have assumed that the zeros scale as
\begin{align}
v_j& \sim -\frac{\log{|t|}}{4}+\frac{\b^{(1)}_j}{5}=\frac{\pi\varepsilon}{2}+\frac{\b^{(1)}_j}{5}\\
w_k& \sim -\frac{\log{|t|}}{4}+\frac{\b^{(2)}_k}{5}=\frac{\pi\varepsilon}{2}+\frac{\b^{(2)}_k}{5}
\end{align}
or, more precisely, the scaled locations of the zeros are defined by
\begin{eqnarray}
\b^{(1)}_j&=&\lim_{N\to\infty,t\to 0}\Big(5v_j+\frac{5}{4}\log |t|\Big)
=\lim_{N\to\infty,t\to 0}\Big(5v_j+\log\frac{\mu}{N}\Big)\\
\b^{(2)}_k&=&\lim_{N\to\infty,t\to 0}\Big(5w_k+\frac{5}{4}\log |t|\Big)
=\lim_{N\to\infty,t\to 0}\Big(5w_k+\log\frac{\mu}{N}\Big)
\end{eqnarray}
Close to the UV fixed point, when $mR$ is small, we must have $\b^{(1)}_j<0$ and $\b^{(2)}_k>0$.

If we define the rapidity  $\vartheta$ and pseudo-energies $\epsilon_i(\vartheta)$ by
\begin{equation}
e^{-\epsilon_i(\vartheta)}=\lim_{N\to\infty} t_i\Big(\vartheta-\log\frac{\mu}{N}\Big)
=\hat{t}_i(\vartheta-\log\mu),\qquad i=1,2
\end{equation}
where $\mu=mR/4$ we obtain the TBA equations 
\begin{eqnarray}
\label{e:TBA}
\u_1(\h)\!\!&=&\!\!mRe^\h -\log{e^\h+\sqrt{2}+e^{-\h}\over e^\h-\sqrt{2}+e^{-\h}}-
\frac{1}{2\pi}\int_{-\infty}^{\infty}\!d\h'\,\frac{\log(1+e^{-\u_2(\h')})}
{\cosh(\h-\h')}\nonumber\\
&&\mbox{}-\sum_{j=1}^{m_1}\log\tanh\big(\frac{\b^{(1)}_j}{2}-\frac{\h}{2}\big)
-\sum_{k=1}^{m_2}\log\tanh\big(\frac{\b^{(2)}_k}{2}-\frac{\h}{2}\big)\nonumber\\[4pt]
\u_2(\h)\!\!& =&\!\! mRe^{-\h}-\log{e^\h+\sqrt{2}+e^{-\h}\over e^\h-\sqrt{2}+e^{-\h}}-
\frac{1}{2\pi}\int_{-\infty}^{\infty}\!d\h'\,\frac{\log(1+e^{-\u_1(\h')})}
{\cosh(\h-\h')}\\
&&\mbox{}-\sum_{j=1}^{m_1}\log\tanh\big(\frac{\b^{(1)}_j}{2}+\frac{\h}{2}\big)
-\sum_{k=1}^{m_2}\log\tanh\big(\frac{\b^{(2)}_k}{2}+\frac{\h}{2}\big)\nonumber
\end{eqnarray}

To find the locations $\b^{(1)}_j$, $\b^{(2)}_k$ of the 1-strings  consider the functional equations
\begin{eqnarray}
t_1(x-i\frac{\pi}{2})t_1(x+i\frac{\pi}{2}) &=& 1 + t_2(x)\nonumber\\
t_2(x-i\frac{\pi}{2})t_2(x+i\frac{\pi}{2}) &=& 1 + t_1(x)
\end{eqnarray}
at $x=-\pi i/2+5v_j$, $x=-\pi i/2+5w_k$ respectively.
Since the right-hand sides vanish,  in the scaling limit this implies
\begin{eqnarray}
\hat{t}_2(\b_j^{(1)}-\frac{\pi i}{2}-\log\mu) = -1 = e^{-n^{(1)}_j\pi i},\quad j=1,2,\ldots,m_1
\nonumber\\
\hat{t}_1(\b_k^{(2)}-\frac{\pi i}{2}-\log\mu) = -1 = e^{-n^{(2)}_k\pi i},\quad k=1,2,\ldots,m_2
\end{eqnarray}
or
\begin{eqnarray}
\u_2(\b^{(1)}_j-{\pi i\over 2})&=&n^{(1)}_j\pi i,\qquad j=1,2,\ldots,m_1\nonumber\\
\u_1(\b^{(2)}_k-{\pi i\over 2})&=&n^{(2)}_k\pi i,\qquad k=1,2,\ldots,m_2
\label{pseudoquantization}
\end{eqnarray}
where $n^{(1)}_j$, $n^{(2)}_k$ are odd integers. These integers are given by their values~\cite{OPW97} in the UV limit as determined by winding numbers
\begin{eqnarray}
n_j^{(1)}& =&
2(m_1-j)-m_2 + 1 + 2I_j^{(1)},\qquad j=1,2,\ldots,m_1\nonumber\\
n_k^{(2)}& =&
2(m_2-k)-m_1 + 1 + 2I_k^{(2)},\qquad k=1,2,\ldots,m_2
\label{nkquantum}
\end{eqnarray}
where $I=(I^{(1)}_1,\ldots,I^{(1)}_{m_1}|I^{(2)}_1,\ldots,I^{(2)}_{m_2})$ are the quantum numbers (\ref{quantumnos}). 
These integers $n^{(i)}_\ell$ can change during the flow due to winding of phases. 
For numerical purposes, a more useful form of these auxiliary equations is obtained by replacing $\vartheta$ with $\b^{(i)}_\ell-\frac{\pi i}{2}$ in the TBA equations (\ref{e:TBA}). 
Similar equations can obtained for the locations of the 2-strings.

Repeating the same calculation as in Regime~III, we find
the finite-size corrections to the scaled energies in Regime~IV are
\begin{equation}
\frac{1}{2}\log{D_1(x)}  =-\frac{RE(R)\cosh{x}}{2N}+o(\frac{1}{N})
\end{equation}
with
\begin{equation}
R E(R) =mR\Big[\!\sum_{j=1}^{m_1}\!\!
e^{-\b^{(1)}_j}\!+\!\sum_{k=1}^{m_2}\! e^{-\b^{(2)}_k}\Big]
-\frac{mR}{2\pi}\!\int_{-\infty}^{\infty}\!d\h\,
e^{\h}\,\log\!{(1\!+\!e^{-\u_1(\h)})}
\label{scaledenergies}
\end{equation}

In the UV limit $m R\to 0$ we recover the critical TBA equations of \cite{OPW97}. 
Explicitly, setting $\vartheta=x+\log\mu$, we find that as  $\mu\to 0$ the location of the 1-strings scale as
\begin{eqnarray}
 \b^{(1)}_j&=&\phantom{-}y^{(1)}_j+\log\mu,\quad j=1,2,\ldots,m_1\nonumber\\
 \b^{(2)}_k&=&-y^{(2)}_k-\log\mu,\quad k=1,2,\ldots,m_2
\end{eqnarray}
It follows that in the limit $mR\to 0$ the finite-size corrections to the scaled energies are given exactly by
\begin{equation}
\frac{1}{2}\log{D}_1(u)=
\frac{2\pi}{N}\sin 5u\Bigl[\frac{7}{240}
-\frac{1}{4}\vm C\vm-\sum_{i=1}^{2}\sum_{j=1}^{m_i}I^{(i)}_j
\Bigr]
\end{equation}
with the finitized partition function
\begin{equation}
Z_N(q) = q^{-7/240}
\sum_{
m_1,\,m_2\,\text{even}}
q^{\tfrac{1}{4}\vm C\vm}
\Mult{m_1+n_1}{m_1}{q}\Mult{m_2+n_2}{m_2}{q}
\end{equation}
where the modular parameter is $q=\exp(-\pi{\sin 5u}M/N)$ for $M$ double rows.

We have chosen to write the TBA equations in terms of both strips~1 and 2. However, in accord with the existence of a single extended strip, the set of TBA equations (\ref{e:TBA}), (\ref{pseudoquantization}), (\ref{nkquantum}) and (\ref{scaledenergies}) can be written in terms of a single strip by using the symmetry $\epsilon_2(\vartheta)=\epsilon_1(-\vartheta)$. Explicitly, extending strip~1, we obtain the TBA equations
\begin{eqnarray}
\label{singleTBA}
\u_1(\h)&=&mRe^\h -\log{e^\h+\sqrt{2}+e^{-\h}\over e^\h-\sqrt{2}+e^{-\h}}-
\frac{1}{2\pi}\int_{-\infty}^{\infty}\!d\h'\,\frac{\log(1+e^{-\u_1(\h')})}
{\cosh(\h+\h')}\nonumber\\
&&\qquad
\mbox{}-\sum_{\ell=1}^m\log\tanh\big(\frac{\b^{(1)}_\ell}{2}-\frac{\h}{2}\big)\\
R E(R)&=&mR\sum_{\ell=1}^m e^{-\b^{(1)}_\ell}
-\frac{mR}{2\pi}\!\int_{-\infty}^{\infty}\!d\h\,
e^{\h}\,\log\!{(1\!+\!e^{-\u_1(\h)})}
\end{eqnarray}
where $m=m_1+m_2$ and
\begin{eqnarray}
\b^{(1)}_{m+1-k}=\b^{(2)}_k,\qquad k=1,2,\ldots,m_2
\end{eqnarray}
The auxiliary equations become
\begin{eqnarray}
\u_1({\pi i\over 2}-\b^{(1)}_\ell)=n^{(1)}_\ell\pi i,\qquad 
\ell=1,2,\ldots,m
\end{eqnarray}
where 
\begin{eqnarray}
n^{(1)}_\ell=m-2\ell+1+2\tilde{I}_\ell,
\qquad\ell=1,2,\ldots,m
\label{singlen}
\end{eqnarray}
and the quantum numbers $\tilde{I}_\ell$ in the extended strip are given by
\begin{eqnarray}
\tilde{I}_\ell=
\begin{cases}
n_2+I^{(1)}_\ell,& \ell=1,2,\ldots,m_1\\
n_2-I^{(1)}_{m+1-\ell},& \ell=m_1+1,m_1+2,\ldots,m
\end{cases}
\end{eqnarray}
with $n_2=m_1/2-m_2$.
This symmetry between strips~1 and 2 is manifestly broken in the UV limit $m R\to 0$ and the IR limit $mR\to\infty$. Also these TBA equations need to be modified in the intermediate regime for Mechanism~A levels after collision of the 1-strings.

\subsection{IR Massless TBA: $(r,s)=(1,1)$}

For large $mR$ we work with the extended strip~1. In this regime the total number of 1-strings is either $m=m_1+m_2$ for Mechanism B, C or $m=m_1+m_2-2$ for Mechanism~A with quantum numbers $I'=(I'_1,I'_2,\ldots,I'_m)$ given by (\ref{Isingquantum}) to (\ref{Isingquantum2}). Setting
\begin{equation}
\epsilon'(\vartheta)=\epsilon_1(\vartheta)+m\pi i=\epsilon_2(-\vartheta)+m\pi i
\label{eps12'}
\end{equation}
we now obtain the single TBA equation
\begin{eqnarray}
\u'(\h)\!\!&=&\!\!mRe^{\h} -\log{e^\h+\sqrt{2}+e^{-\h}\over e^\h-\sqrt{2}+e^{-\h}}-
\frac{1}{2\pi}\int_{-\infty}^{\infty}\!d\h'\,\frac{\log(1+e^{-\u'(\h')})}
{\cosh(\h+\h')}\nonumber\\
&&\mbox{}\qquad-\sum_{\ell=1}^{m}\log\tanh\big(\frac{\b'_\ell}{2}-\frac{\h}{2}\big)
\end{eqnarray}
with scaling energies
\begin{equation}
R E(R) =mR \sum_{\ell=1}^{m} e^{-\b'_\ell}
-\frac{mR}{2\pi}\!\int_{-\infty}^{\infty}\!d\h\,
e^{\h}\,\log{(1+e^{-\u'(\h)})}
\end{equation}
The auxiliary equations are
\begin{eqnarray}
\u'({\pi i\over 2}-\b'_\ell)&=&n'_\ell\pi i,\qquad \ell=1,2,\ldots,m
\label{IsingAux}
\end{eqnarray}
where
\begin{eqnarray}
n_\ell'& =&2(m-\ell) + 1 + 2I_\ell',\qquad \ell=1,2,\ldots,m
\end{eqnarray}
The difference between this equation and (\ref{singlen}),  as reflected in (\ref{eps12'}), arises because of the $m\pi i$ winding between the previous position of the scaling edge at $\pi i\varepsilon/2$ (reference point for $\vartheta=0$) and the new position of the scaling edge at $\pi i\varepsilon$.

Setting $\vartheta=x+\log\mu$ in the IR limit $mR\to\infty$, the locations of the 1-strings scale as
\begin{equation}
\b_\ell'=y_\ell'+\log\mu,\qquad \ell=1,2,\ldots,m
\end{equation}
and the above equations reduce to the TBA equations of the critical Ising model.
Indeed, the pseudo energies $\u_i(\h)$
decouple giving the energy of the usual massless free fermions
\begin{equation}
\u'(\h) \sim mRe^{-\h}
\end{equation}
so that the auxilliary equations \eqref{IsingAux} become 
\begin{equation}
4e^{-y'_\ell} =- [2(m-\ell)+1+2I_\ell']\pi
\end{equation}
In this limit the finite size corrections for the scaled energies are
\begin{equation}
\frac{1}{2}\log{D}_1(x)
=-\frac{\pi\cosh x}{N}\Bigl[
-\frac{1}{48}+E'\Bigr],\qquad E'=\frac{m^2}{2}+\sum_{\ell} I_\ell'
\end{equation}
with the partition function
\begin{equation}
Z_{1,1}^{Ising}(q) = q^{-1/48}
\sum_{m\,even} q^{\frac{m^2}{2}} \Mult{\frac{N+m-2}{2}}{m}{q}.
\end{equation}

\section{Massless Numerics}

The TBA equations of the previous section can be solved numerically by an iterative procedure. There are, however, some subtleties. The process starts with initial guesses for the pseudoenergies $\u_i(\vartheta)$ and 1-string locations, close to the UV or IR fixed points. The
flow is followed by progressively incrementing or decrementing 
$mR$. At each value of $mR$, the TBA equations are used to update the
pseudoenergies $\u_i(\vartheta)$ and then these are used in the 
auxiliary equations to update the locations of the 1-strings, and so on, until a stable solution is reached.  Typically, the UV form of the equations is stable for small values of $mR$, the IR form is stable for large values of $mR$ and there is an intermediate range of values of $mR$ for which both forms are stable and converge (with a precision of five decimals places) to the same values for the scaled energies and the locations of each of the 1-strings. 
In all cases these numerical flows confirm the three mechanisms A, B, C.

One numerical difficulty is related to the
determination of the location of the 1-strings in strip~2. This problem arises because in the UV form of the equations $\b_{k}^{(2)}$ cannot be
obtained by direct iteration of the auxiliary equation. This problem is solved
by inverting the phases 
\begin{equation}
\log\tanh\big(\frac{\b^{(1)}_j-\b^{(2)}_k}{2}+\frac{\pi i}{4}\big)
\end{equation}
A more serious problem relates to Mechanism~A cases for which two 1-strings collide to form a (short) 2-string with complex coordinates 
\begin{equation}
\b^{(1)}_{m_1}, \b^{(2)}_{m_2}\mapsto \b\pm i\gamma
\end{equation}
For these cases there is an intermediate range of values of $mR$ which requires a modification of the TBA equations. Such short 2-strings were studied~\cite{BLZ} in the context of the Yang-Lee scaling theory. Unfortunately, due to instabilities in our equations, we have been unable to numerically solve the Mechanism~A equations throughout the intermediate region.

\begin{figure}[t]
\hskip 1.25truein
\mbox{}\hspace{-.75in}\includegraphics[width=0.85\linewidth]{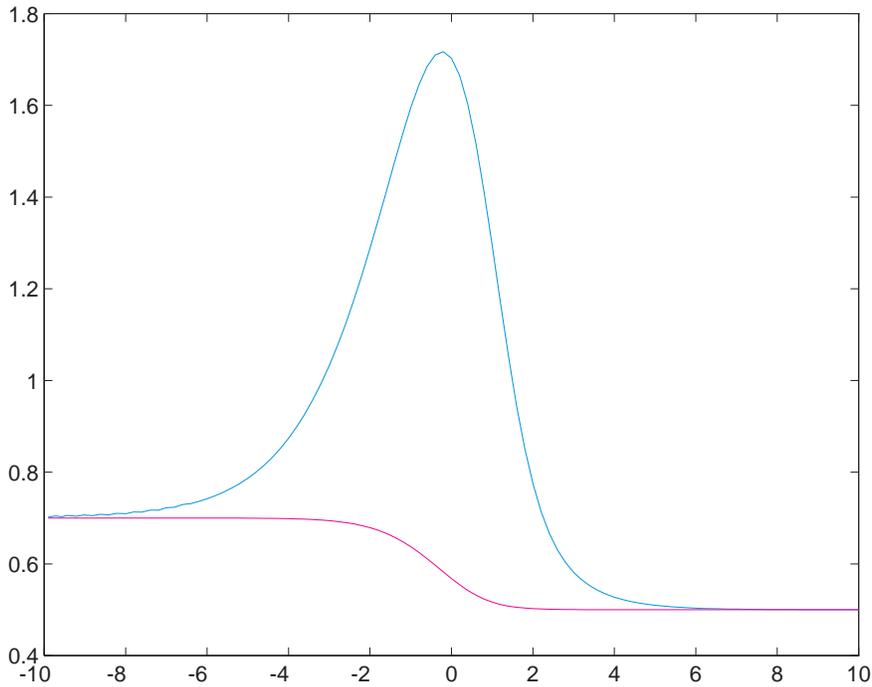}
\vspace{-.25in}
    \caption{Flow of groundstate energy $c_{\mbox{\small eff}}=-12RE(R)/\pi$ versus $\log(mR)$ for periodic and $(r,s)=(1,1)$ boundary conditions. 
The difference arises from the boundary term in the TBA equations. 
The energy is a decreasing function of $mR$ for periodic boundary conditions by Zamolodchikov's $c$-theorem but this theorem does not apply with fixed boundary conditions.}\label{cflow}     
\end{figure}

\begin{figure}[p]
\hskip 1.25truein \vskip -1.75truein
\mbox{}\hspace{-2.5in}
\includegraphics[width=1.75\linewidth]{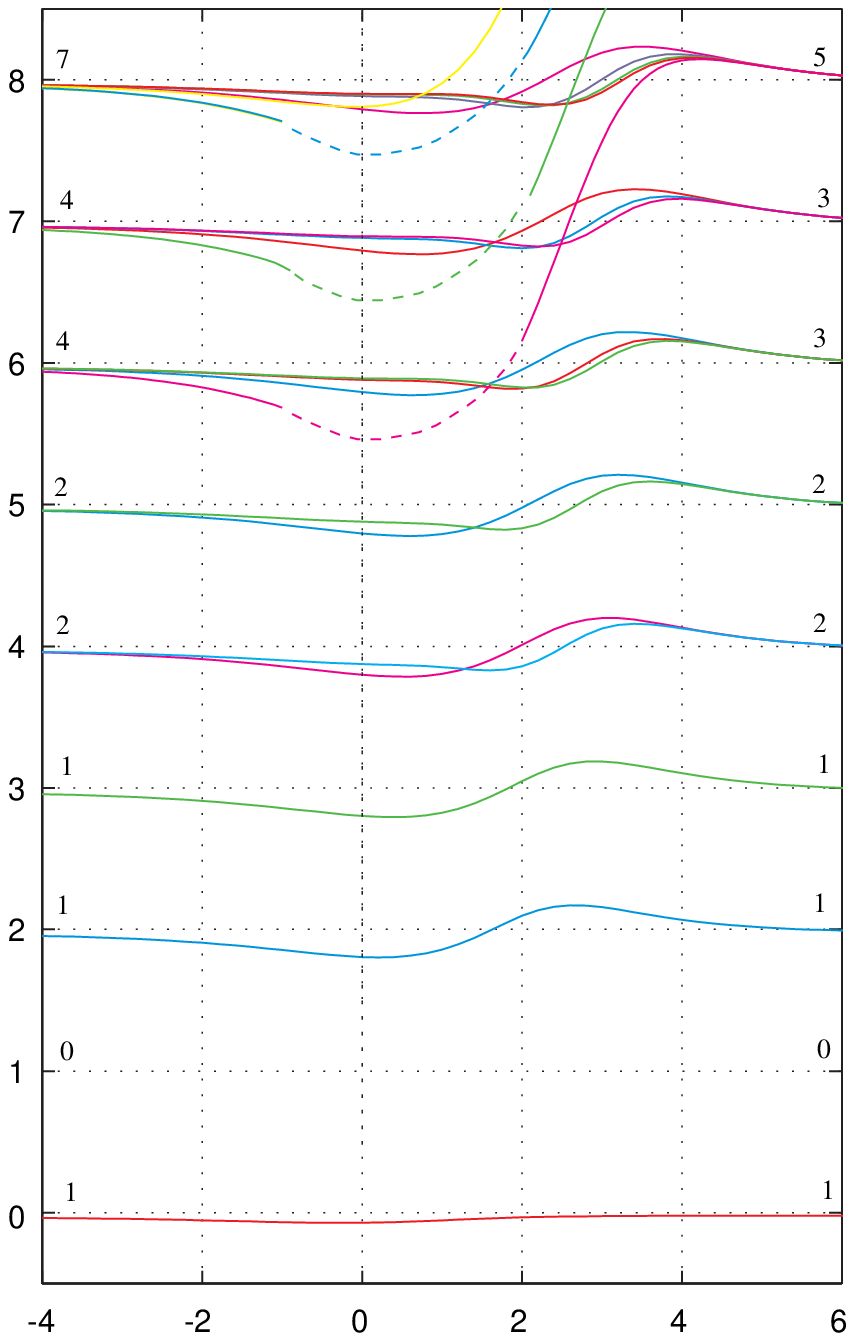}
\vspace{-.75in}
    \caption{Flow $\chi_{1,1}^4(q)\mapsto\chi_{1,1}^3(q)$ of scaling energies $-c_{\mbox{\small eff}}/24=RE(R)/2\pi$ versus $\log(mR)$ in the $(r,s)=(1,1)$ sector. The degeneracies of the levels are shown in the margins. 
The intermediate region of the Mechanism~A levels (shown dashed) are schematic and have not been obtained from numerical solution of the TBA equations.}\label{flow11}     
\end{figure}

\begin{figure}[p]
\vspace{-0.5in}
\hskip 1.25truein
\mbox{}\hspace{-.75in}\includegraphics[width=0.85\linewidth]{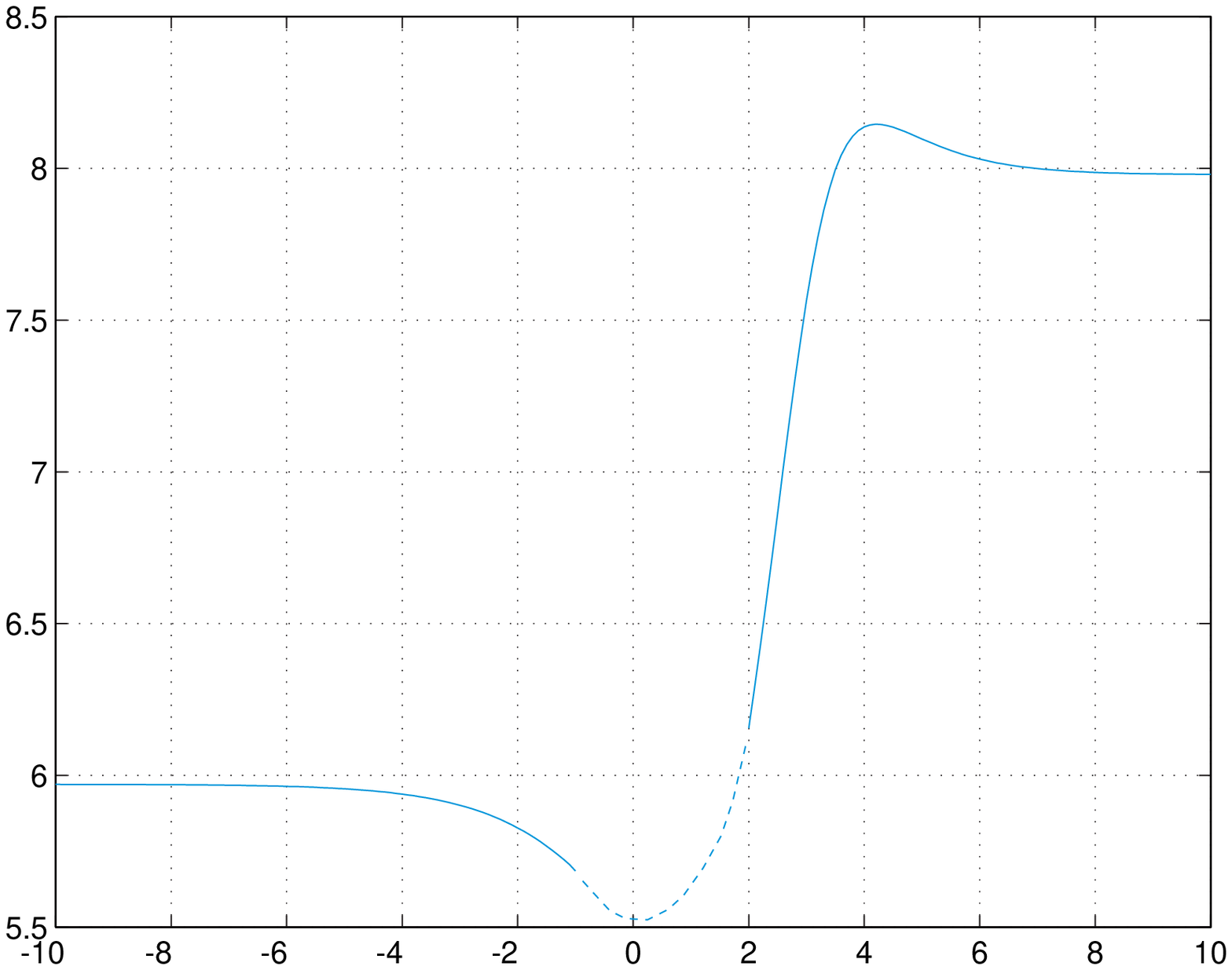}
\vspace{-.25in}
    \caption{The scaling energy $-c_{\mbox{\small eff}}/24=RE(R)/2\pi$ versus $\log(mR)$ for the Mechanism~A level in the $(r,s)=(1,1)$ sector with string contents $(m_1,m_2)=(4,2)$ and UV quantum numbers $I=(0,0,0,0|0,0)$.}    
\hskip 1.25truein
\mbox{}\hspace{-.75in}\includegraphics[width=0.85\linewidth]{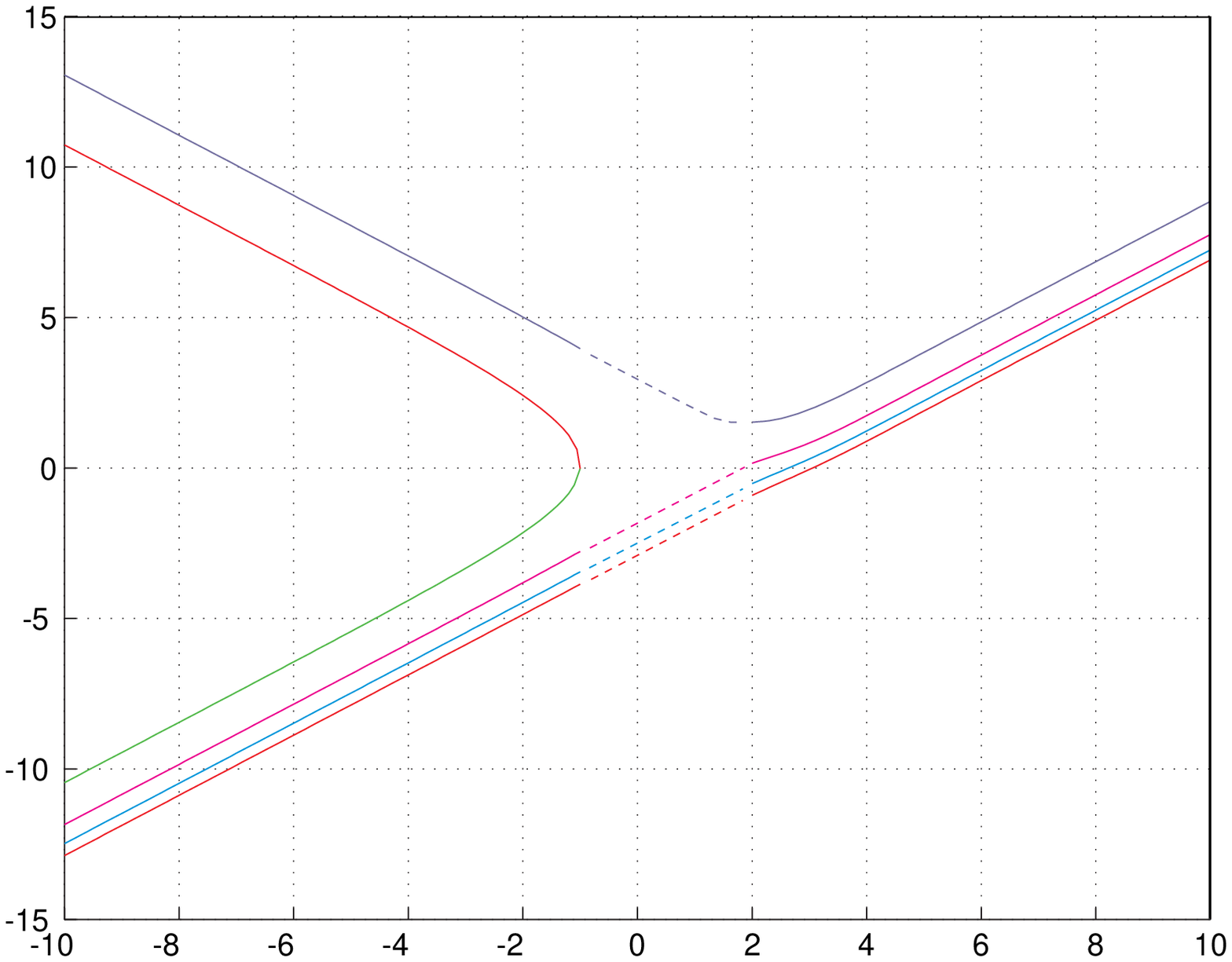}
\vspace{-.25in}
    \caption{Flow of six 1-strings versus $\log(mR)$ for the Mechanism~A level in the $(r,s)=(1,1)$ sector with string contents $(m_1,m_2)=(4,2)$ and UV quantum numbers $I=(0,0,0,0|0,0)$.}     
\end{figure}

\begin{figure}[p]
\vspace{-0.5in}
\hskip 1.25truein
\mbox{}\hspace{-.75in}\includegraphics[width=0.85\linewidth]{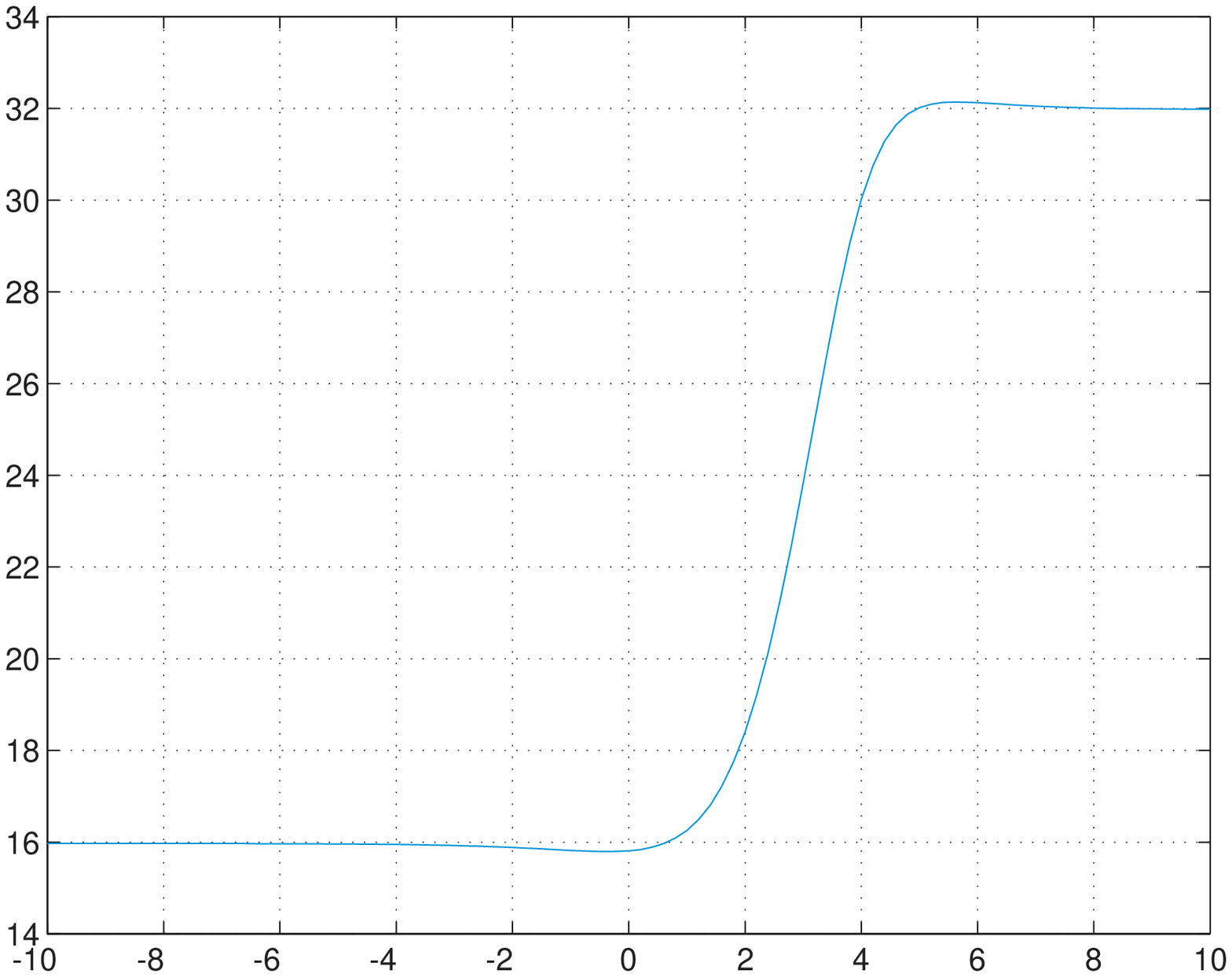}
\vspace{-.25in}
    \caption{The scaling energy $-c_{\mbox{\small eff}}/24=RE(R)/2\pi$ versus $\log(mR)$ for the Mechanism~B level in the $(r,s)=(1,1)$ sector with string contents $(m_1,m_2)=(6,2)$ and UV quantum numbers $I=(0,0,0,0,0,0|1,1)$.}   
\hskip 1.25truein
\mbox{}\hspace{-.75in}\includegraphics[width=0.85\linewidth]{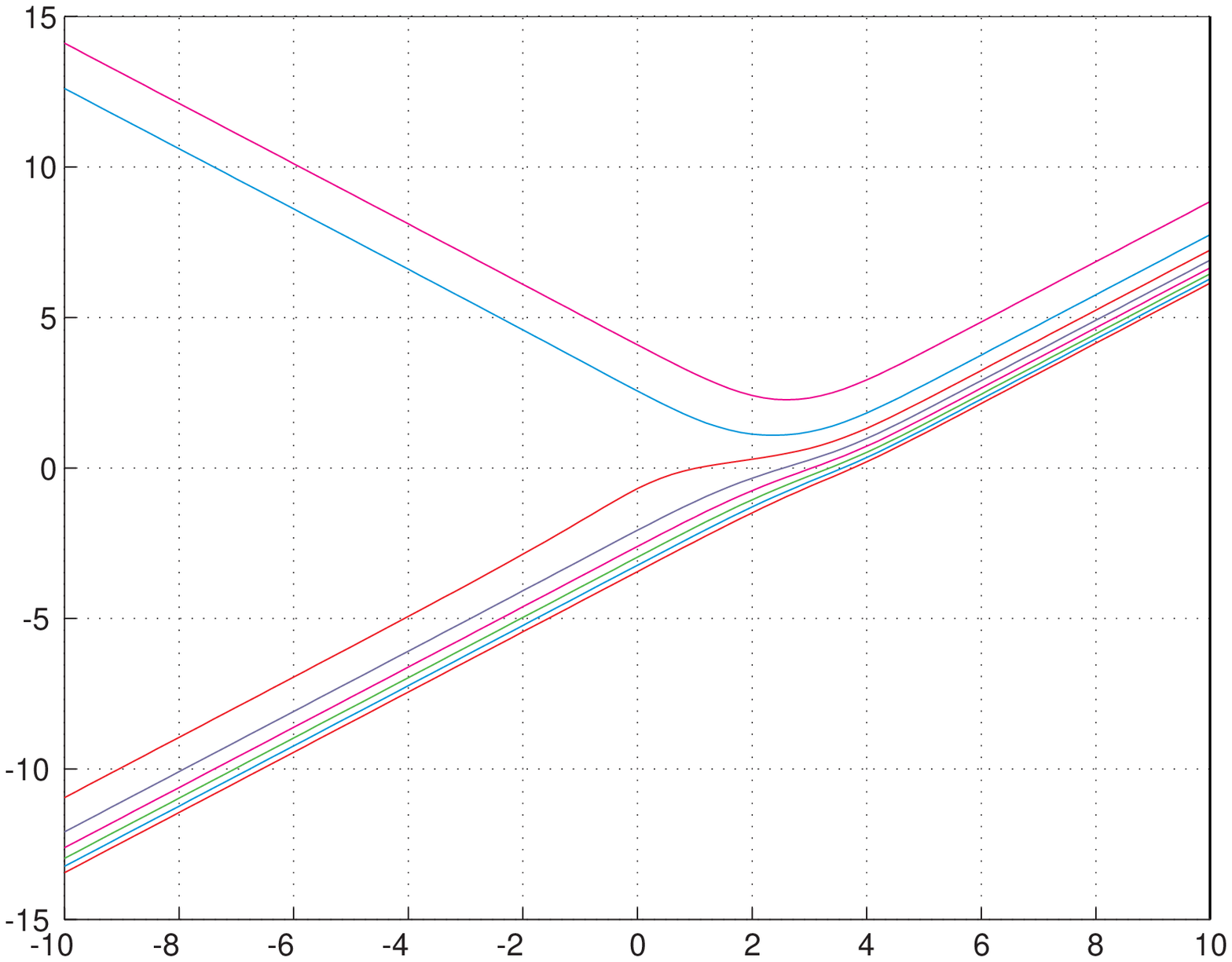}
\vspace{-.25in}
    \caption{Flow of eight 1-strings versus $\log(mR)$ for the Mechanism~B level in the $(r,s)=(1,1)$ sector with string contents $(m_1,m_2)=(6,2)$ and UV quantum numbers $I=(0,0,0,0,0,0|1,1)$.}    
\end{figure}

\begin{figure}[p]
\hskip 1.25truein \vskip -0.75truein
\mbox{}\hspace{-2.5in}
\includegraphics[width=1.75\linewidth]{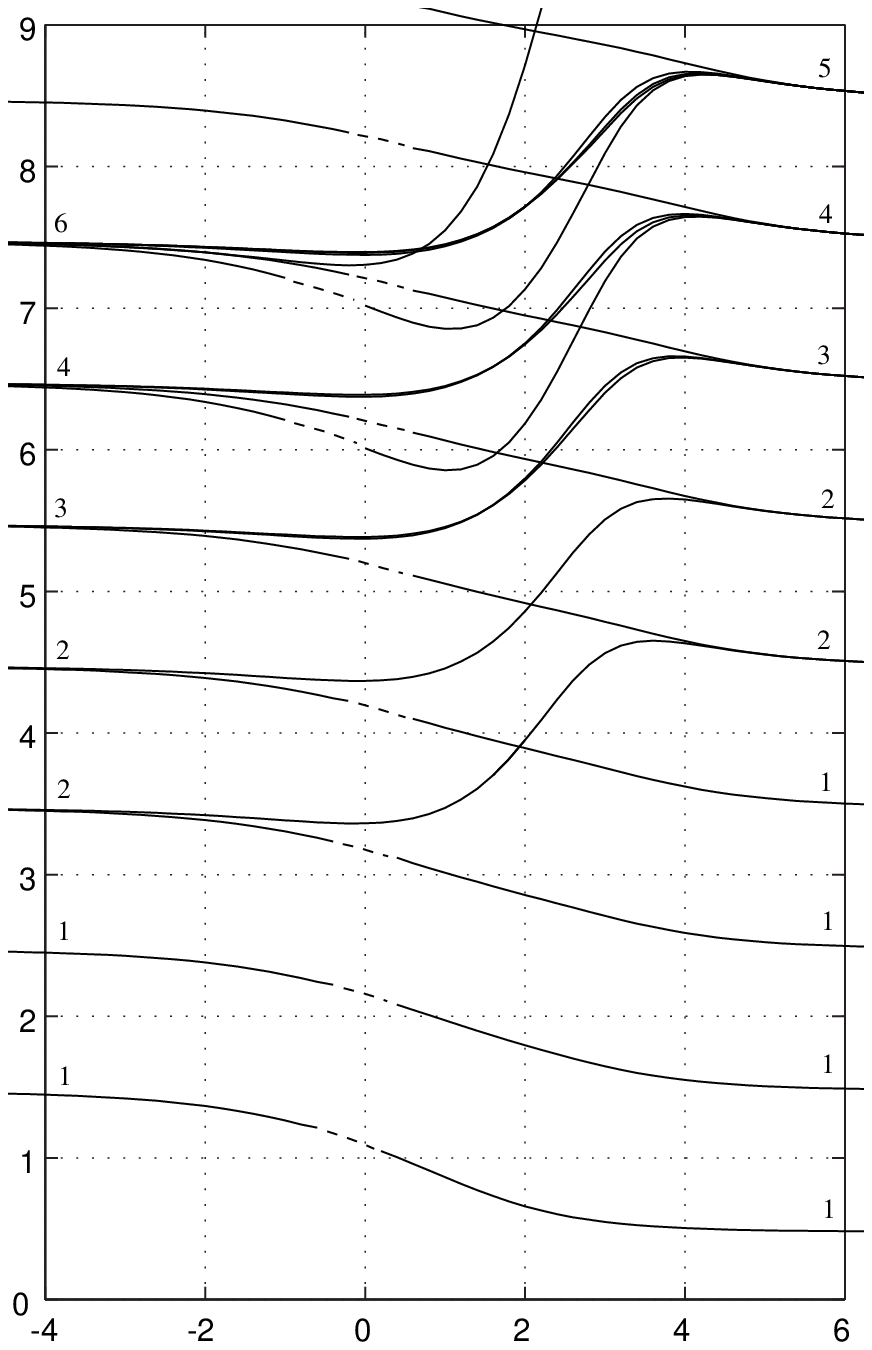}
\vspace{-.75in}
    \caption{Flow $\chi_{3,1}^4(q)\mapsto\chi_{1,3}^3(q)$ of scaling energies $-c_{\mbox{\small eff}}/24=RE(R)/2\pi$ versus $\log(mR)$ in the $(r,s)=(3,1)$ sector. The degeneracies of the levels are shown in the margins. 
The intermediate region of the Mechanism~A levels (shown dashed) are schematic and have not been obtained from numerical solution of the TBA equations.}\label{flow31}     
\end{figure}

We present some typical numerical results in a series of figures. In Figure~3, we compare the groundstate scaling energy in the $(r,s)=(1,1)$ sector with the scaling energy for periodic boundary conditions. In Figures~4 and 9 we show 
the flow of scaling energies in the sectors $(r,s)=(1,1)$ and $(r,s)=(3,1)$ respectively. A dashed curve is used to guide the eye in the intermediate regime of the Mechanism~A level. 
In Figures~5 and 6, we show the flow of the scaling energy and 1-strings for the Mechanism~A level in this sector with string contents $(m_1,m_2)=(4,2)$ and  UV quantum numbers $I=(0,0,0,0|0,0)$. The dashed curves in the intermediate regime are schematic and have not been calculated from the solution of the TBA equations. 
For comparison, we show  in Figures~7 and 8, the flow of the scaling energy and 1-strings for the Mechanism~B level with string contents $(m_1,m_2)=(6,2)$ and quantum numbers $I=(0,0,0,0,0,0|1,1)$. The flow of the scaling energies and 1-strings for arbitrary Mechanism~B and C levels can be calculated throughout the flow by numerical solution of the TBA equations. The mechanism~A levels can be calculated right up to the point where the two 1-strings collide. 
Note the linear regimes in the UV and IR for the flows of 1-strings. This corresponds to the assumed limiting scaling of the locations of these 1-strings.

\section{Discussion}

In this paper we have used a lattice approach to derive TBA equations for all excitations in the massless renormalization group flow from the tricritical to critical Ising model. The excitations are classified according to string content which changes by one of three mechanisms A,B,C along the flow and leads to a mapping between finitized Virasoro characters. With the exception of the intermediate regime for Mechanism~A flows, the TBA equations can be solved numerically by iteration. It would be of interest to compare our results with the results of the Truncated Conformal Space Approximation.

Although, the tricritical Ising model is superconformal, the boundary conditions applied in this paper break the superconformal symmetry. It would be of interest to investigate the pattern of the superconformal flows between fixed points corresponding to superconformal boundary conditions~\cite{RichardP,Nepo02}. It would also be of interest to extend the analysis of this paper to the complete flow for periodic boundary conditions.

\clearpage
\section*{Appendix}
\setcounter{section}{1}
\renewcommand{\theequation}{\Alph{section}.\arabic{equation}}

In this Appendix we prove the generalized $q$-Vandermonde identities of Section~2.4. We simplify
the identities to show they are special cases of general identities obtained by
Bender~\cite{Bender71}. We then give an elementary proof of these identities using induction. 

To simplify the identities of Section~2.4 we set $m_1=2k$ or $m_1=2k+1$
depending on the parity of $m_1$ and set $n=(N+m)/2$ or $n=(N+m\pm1)/2$ as appropriate. This
reduces the six identities to four identities
\begin{eqnarray}
&&\chi_{1,1}/\chi_{3,1}:\qquad\gauss{n}{m}=\sum_k q^{6k^2-2km}\left\{
               \gauss{n-k}{2k}     \gauss{k}{m-2k}\right.\\ 
&&\left.\mbox{}\hspace{.5in}+
q^{-2k}        \gauss{n-k}{2k-1}   \gauss{k-1}{m-2k}+
q^{m-7k+2}		\gauss{n-k+1}{2k-1} \gauss{k-1}{m-2k+1}\right\}\nonumber\\
&&\chi_{2,2}:\qquad\qquad\gauss{n}{m}=\sum_{k} q^{6k^2-2km}\left\{
				           \gauss{n-k}{2k}   \gauss{k}{m-2k}\right.\\ &&\left.\mbox{}+
q^{-m+5k+1} \gauss{n-k-1}{2k} \gauss{k}{m-2k-1}+
q^{m-5k+1}  \gauss{n-k}{2k-1} \gauss{k}{m-2k+1}\right\}\nonumber\\
&&\chi_{2,1}:\qquad\qquad\gauss{n}{m}=\sum_{k} q^{6k^2-2km-m+6k+3/2}\left\{
q^{k+1/2}		      \gauss{n-k-1}{2k+1}    \gauss{k+1}{m-2k-1}\right.\\ &&\left.\mbox{}+
q^{-k-1/2}       \gauss{n-k-1}{2k}      \gauss{k}{m-2k-1}+
q^{m-6k-3/2}   \gauss{n-k}{2k}        \gauss{k}{m-2k}\right\}\nonumber\\
&&\chi_{1,2}/\chi_{3,2}:\qquad\gauss{n}{m}=\sum_{k} q^{6k^2-2km-m+6k+3/2}\left\{
q^{m-6k-3/2}		 \gauss{n-k}{2k}  \gauss{k}{m-2k}\right.\\ &&\left.\mbox{}+
q^{-m+4k+5/2}  \gauss{n-k-1}{2k+1}  \gauss{k}{m-2k-2}+
q^{-k-1/2}       \gauss{n-k}{2k+1}    \gauss{k}{m-2k-1}\right\}\nonumber
\end{eqnarray}
After some recasting, a surprising mod~3 property emerges in the terms of these identities
\begin{eqnarray}
&&\chi_{1,1}/\chi_{3,1}:\qquad\gauss{n}{m}=\sum_k q^{6k^2-2km}\left\{
               \gauss{n-k}{n-3k}     \gauss{k}{3k-m}\right.\\ &&\left.\mbox{}+
q^{-2k}        \gauss{n-k}{n-3k+1}   \gauss{k-1}{3k-1-m}+
q^{m-7k+2}		\gauss{n-k+1}{n-3k+2}    \gauss{k-1}{3k-2-m}\right\}\nonumber\\
&&\chi_{2,2}:\qquad\qquad\gauss{n}{m}=\sum_{k} q^{6k^2-2km}\left\{
				           \gauss{n-k}{n-3k}   \gauss{k}{3k-m}\right.\\ &&\left.\mbox{}+
q^{-m+5k+1} \gauss{n-k-1}{n-3k-1} \gauss{k}{3k+1-m}+
q^{m-5k+1}  \gauss{n-k}{n-3k+1} \gauss{k}{3k-1-m}\right\}\nonumber
\end{eqnarray}
\begin{eqnarray}
&&\chi_{2,1}:\qquad\qquad\gauss{n}{m}=\sum_{k} q^{6k^2-2km}\left\{
q^{-m+7k+2}		      \gauss{n-k-1}{n-3k-2}    \gauss{k+1}{3k+2-m}\right.\\ &&\left.\mbox{}+
q^{-m+5k+1}       \gauss{n-k-1}{n-3k-1}      \gauss{k}{3k+1-m}+
                   \gauss{n-k}{n-3k}        \gauss{k}{3k-m}\right\}\nonumber\\
&&\chi_{1,2}/\chi_{3,2}:\qquad\qquad\gauss{n}{m}=\sum_{k} q^{6k^2-2km}\left\{
                   		 \gauss{n-k}{n-3k}  \gauss{k}{3k-m}\right.\\ &&\left.\mbox{}+
q^{-2m+10k+4}    \gauss{n-k-1}{n-3k-2}  \gauss{k}{3k+2-m}+
q^{-m+5k+1}       \gauss{n-k}{n-3k-1}    \gauss{k}{3k+1-m}\right\}\nonumber
\end{eqnarray}
Setting $\ell=3k,3k+1,3k+2$ mod~3 reduces the four identities to just two identities
\begin{eqnarray}
&&\mbox{}\hspace{-.75in}
\chi_{1,1}/\chi_{3,1}/\chi_{1,2}/\chi_{3,2}:\qquad\gauss{n}{m}=\sum_\ell
q^{(\ell-\floor{\ell/3})(\ell-m)}
               \gauss{n-\floor{(\ell+1)/3}}{n-\ell}     \gauss{\floor{\ell/3}}{\ell-m}\\
&&\mbox{}\hspace{-.75in}
\chi_{2,2}/\chi_{2,1}:\qquad\qquad\qquad\;
															\gauss{n}{m}=\sum_\ell q^{(\ell-\floor{(\ell+1)/3})(\ell-m)}
               \gauss{n-\floor{(\ell+2)/3}}{n-\ell}     \gauss{\floor{(\ell+1)/3}}{\ell-m}
\end{eqnarray}
For $(n,m)\ne(0,0)$ there is an additional identity
\begin{equation}
\gauss{n}{m}=\sum_\ell q^{(\ell-\floor{(\ell+2)/3})(\ell-m)}
               \gauss{n-\floor{(\ell+3)/3}}{n-\ell}     \gauss{\floor{(\ell+2)/3}}{\ell-m}
\end{equation}
These are in fact special cases\footnote{We thank George Andrews for pointing this out to us.} of
identities due to Bender~\cite{Bender71}.

\medskip
\noindent
{\bf Generalized $q$-Vandermonde Identities:}
\begin{equation}
\gauss{n}{m}=\sum_\ell q^{\floor{(2\ell+2-a)/3}(\ell-m)}
               \gauss{n-\floor{(\ell+1+a)/3}}{n-\ell}     \gauss{\floor{(\ell+a)/3}}{\ell-m},
\quad -m\le a\le 2n+1
\end{equation}
{\it Proof:\ } For $n=m$ we have $\ell=m$ so $\mbox{LHS}=\mbox{RHS}=1$ for $-m\le a\le 2m+1$. 
We now proceed by induction on $n$. Suppose that $-m\le a\le 2n+1$ and $-(m-1)\le a+1\le 2n+1$, that is,
$-m\le a\le 2n$. Then
\begin{eqnarray*}
\gauss{n+1}{m}&=&\gauss{n}{m}+q^{n-m+1}\gauss{n}{m-1}
=\gauss{n}{m}+q^{(n-\ell)+(\ell-m+1)}\gauss{n}{m-1}\\
&=&\sum_\ell q^{\floor{(2\ell+2-a)/3}(\ell-m)}
               \gauss{n-\floor{(\ell+1+a)/3}}{n-\ell}     \gauss{\floor{(\ell+a)/3}}{\ell-m}\\
&&\qquad\mbox{}+\sum_\ell q^{(\floor{(2\ell+1-a)/3}+1)(\ell-m+1)}
      q^{n-\ell} \gauss{n-\floor{(\ell+2+a)/3}}{n-\ell}     \gauss{\floor{(\ell+1+a)/3}}{\ell-m+1}\\
&=&\sum_\ell q^{\floor{(2\ell+2-a)/3}(\ell-m)}
   \left\{ \gauss{n-\floor{(\ell+1+a)/3}}{n-\ell}+q^{n-\ell+1}\gauss{n-\floor{(\ell+1+a)/3}}{n-\ell+1}\right\}
              \gauss{\floor{(\ell+a)/3}}{\ell-m}\\
&=&\sum_\ell q^{\floor{(2\ell+2-a)/3}(\ell-m)}
              \gauss{n+1-\floor{(\ell+1+a)/3}}{n+1-\ell}
              \gauss{\floor{(\ell+a)/3}}{\ell-m}
\end{eqnarray*}
Now suppose that $a=2n+1$. Then only the terms $\ell=n$ and $\ell=n+1$ survive on the RHS and
\begin{eqnarray*}
&&\sum_\ell q^{\floor{(2\ell+2-a)/3}(\ell-m)}
              \gauss{n+1-\floor{(\ell+1+a)/3}}{n+1-\ell}
              \gauss{\floor{(\ell+a)/3}}{\ell-m}\\
&=&\sum_\ell q^{\floor{(2\ell-2n+1)/3}(\ell-m)}
              \gauss{n+1-\floor{(\ell+2n+2)/3}}{n+1-\ell}
              \gauss{\floor{(\ell+2n+1)/3}}{\ell-m}\\
&=&\gauss{n}{m}+q^{n-m+1}\gauss{n}{m-1}=\gauss{n+1}{m}
\end{eqnarray*}


\section*{Acknowledgements}

PAP is supported by the Australian Research Council and thanks the Asia Pacific Center for Theoretical Physics for support to visit Seoul.  CA is supported
in part by Korea Research Foundation 2002-070-C00025, KOSEF 1999-00018. We thank Giuseppe Mussardo for discussions.

\end{document}